\documentclass[english]{article}
\pdfoutput=1
\usepackage[T1]{fontenc}
\usepackage[latin9]{inputenc}
\usepackage{geometry}
\usepackage{amsmath}
\usepackage{amssymb}
\usepackage{graphicx}

\usepackage{ae,aecompl}
\usepackage{lmodern}

\usepackage{color}
\definecolor{darkblue}{rgb}{0,0,0.6}
\definecolor{darkred}{rgb}{0.6,0,0}
\definecolor{darkgrey}{rgb}{0.6,0.6,0.6}

\usepackage[colorlinks=true,urlcolor=darkblue,citecolor=darkblue,linkcolor=darkred,hyperfootnotes=false]{hyperref}

\newcommand{\esc}{\text{esc}}
\newcommand{\cri}{\text{cr}}
\newcommand{\sa}{\text{s}}
\newcommand{\bl}{\text{bl}}
\newcommand{\bu}{\text{B}}
\newcommand{\dd}{\text{d}}

\newenvironment{casesb}
    {
     \arraycolsep=1.1pt
     \left\{
       \begin{array}{rl}
    }
    { 
       \end{array}
     \right.
    }

\DeclareFontFamily{U}{mathb}{\hyphenchar\font45}
\DeclareFontShape{U}{mathb}{m}{n}{
      <5> <6> <7> <8> <9> <10> gen * mathb
      <10.95> mathb10 <12> <14.4> <17.28> <20.74> <24.88> mathb12
      }{}
\DeclareSymbolFont{mathb}{U}{mathb}{m}{n}
\DeclareFontSubstitution{U}{mathb}{m}{n}

\let\dot\relax
\DeclareMathAccent{\dot}{0}{mathb}{"39}
\let\ddot\relax
\DeclareMathAccent{\ddot}{0}{mathb}{"3A}
\let\dddot\relax
\DeclareMathAccent{\dddot}{0}{mathb}{"3B}
\let\ddddot\relax
\DeclareMathAccent{\ddddot}{0}{mathb}{"3C}

\makeatletter
\date{}

\makeatother

\usepackage{babel}
\begin{document}
\title{Nonlocal stationary probability distributions and escape rates for an active Ornstein--Uhlenbeck particle}
\maketitle

\vspace*{-10mm}
\begin{center}
\large
Eric Woillez$^{1}$,  Yariv Kafri$^{1}$, Vivien Lecomte$^{2}$
\end{center}

\smallskip
\begin{center}
{$^1$Department of Physics, Technion, Haifa 32000, Israel}
\\[1mm]
{$^2$Universit\'e Grenoble Alpes, CNRS, LIPhy, F-38000 Grenoble, France}
\end{center}

\bigskip

\begin{abstract}
We evaluate the steady-state distribution and escape rate for an Active Ornstein--Uhlenbeck Particle (AOUP) using methods from the theory of large deviations. The calculation is carried out both for small and large memory times of the active force in one-dimension. We compare our results to those obtained in the literature about colored noise processes, and we emphasize their relevance for the field of active matter. In particular, we stress that contrary to equilibrium particles, the invariant measure of such an active particle is a non-local function of the potential.
This fact has many interesting consequences, which we illustrate through two phenomena.  First, active particles in the presence of an asymmetric barrier tend to accumulate on one side of the potential --~a ratchet effect that was missing is previous treatments. Second, an active particle can escape over a deep metastable state without spending any time at its bottom.
\end{abstract}

\tableofcontents
\newpage

\section{Introduction}

The Active Ornstein--Uhlenbeck Process (AOUP in the following) is defined in one dimension
through the equations of motion
\begin{equation}
\begin{casesb}
\dot{x} & =v-U'(x)\,,\\
\tau\dot{v} & =-v+\sqrt{2D}\xi(t)\,.
\end{casesb}\label{eq:AOUP model}
\end{equation}
The model can be interpreted as describing the overdamped
dynamics of a particle with position $x$ in an external potential
$U$ (with the mobility set to one) and driven by a Gaussian noisy
velocity $v$ with zero average and a time-time correlation function $\left\langle v(t)v(t')\right\rangle =\frac{D}{\tau}e^{-\left|t-t'\right|/\tau}$. First studied in the context
of Langevin equations with memory~\cite{pesquera_path_1983,luciani_functional_1987,klosek-dygas_colored_1988,bray_instanton_1989,hanggi_path_1989,mckane_path_1990,bray_path_1990,luckock_path_1990,jung1987dynamical,hanggi1995colored},
it has received much attention recently when it became one of the
canonical models used to describe the motion of active particles~\cite{szamel2014self,koumakis2014,Farage2015,szamel2015glassy,fodor2016far,caprini2019active}.
Active particles are a particular class of systems where each
particle consumes energy from its environment in order to self-propel. This AOUP model captures this through the stochastic
active force $v$ which has a persistence time $\tau$ that breaks time
reversal symmetry, and therefore drives the system out of equilibrium:
when $\tau$ goes to zero, one recovers the model of equilibrium Brownian
particles, but as long as $\tau>0$, there is a nonzero entropy production
(of order $\tau^{2}$ as $\tau\to 0$~\cite{fodor2016far}).

The model can be solved exactly when $U$ is a harmonic potential. Then, it can be shown that the AOUP process
can be mapped to an equilibrium process. The invariant measure in the potential and the escape rate, assuming that the particle escapes when it reaches a given potential height, can be computed exactly (see for example,~\cite{woillez2019active}). For general potentials most of the work has focused on the limit $\tau\rightarrow0$. There, approximation techniques, specifically the Unified Colored Noise approximation (UCNA) and Fox approximation, have been employed in order to obtain the steady-state distribution
\cite{jung1987dynamical,hanggi1995colored}. These show that to first
order in $\tau$, the AOUP is similar to an equilibrium
system but with an effective {\it local} potential whose form can be obtained and which depends on the second spatial derivative of $U$~\cite{Farage2015}. The term local emphasizes the fact that the steady-state distribution is a local function of the potential and its derivatives. The
result, however, is restricted to convex potentials, with the additional constraint
that the ratio $\sqrt{\left\langle v^{2}\right\rangle }=\sqrt{D/\tau}$
remains large compared to typical values of the force $U'$. To this end, bistable potentials which cannot be treated using the UCNA and Fox approximations, have been studied numerically in~\cite{caprini2019active} where an alternative approximation scheme in the small $\tau$ limit was developed. 

 More recently the escape problem of an AOUP from a general trap was considered in the large $\tau$ limit~\cite{woillez2019active}. In this regime, it was shown that the behavior changes in a drastic manner from the leading order in $\tau$ result at small $\tau$. Specifically, the escape rate from the trap does not depend on the maximum difference in the value of an effective potential, but on the {\it maximal force} exerted on the particle inside the trap. This suggests that the steady-state distribution takes on a very different form when orders higher than $\tau$ are accounted for. The following paper pursues this line of thought by calculating both the steady-state distribution and the closely related problem of the escape rate from a trap, in the small $D$ limit, both to order $\tau^2$ and in the limit of $\tau \to \infty$. We show that in both limits the physics becomes completely different from that revealed by the leading order in $\tau$ result. In particular, the steady-state distribution is now a {\it non-local} function of the potential $U$. This is crucial since, as we illustrate, fundamental non-equilibrium physical processes such as ratchet effects are only displayed when the non-locality of the steady state appears. These, for example in the seminal work of~\cite{lambert2010collective,di2010bacterial,Angelani2011}, are considered one of the hallmarks of active matter.
 
 To obtain these results, as stated above and in contrast to previous studies
of the AOUP model, we focus on the small-noise limit. Namely, we consider the problem when $D$, the amplitude of the noise which drives the variations in $v$, is small. The use of a small noise asymptotic analysis to study the problem has been first applied about three decades ago in the context of Brownian motion subject to colored noise%
\footnote{%
In an earlier version of the paper we were not aware of these pioneering works. We thank an anonymous referee for bringing our attention to Refs.~\cite{bray_instanton_1989,mckane_path_1990,bray_path_1990} and apologize to their authors.}\!. In particular, a path integral formulation was derived, despite the non-Markovian evolution of the position~\cite{pesquera_path_1983,luciani_functional_1987,wio_path-integral_1989,bray_instanton_1989,hanggi_path_1989,mckane_path_1990}. These formulations in turn allow one to study the $D\to 0$ asymptotic regime through an optimal path approach. In particular, they allow for the computation of escape rates~\cite{wio_path-integral_1989,bray_instanton_1989,bray_path_1990} including the sub-dominant prefactor~\cite{luckock_path_1990} and even accounting for inertia~\cite{newman_inertial_1990} (see also~\cite{luciani_functional_1987} for solutions of specific models or~\cite{mckane_noise-induced_1989} for other correlated noise). 
We also note an early work by K{\l}osek-Dygas, Matkowsky and Schuss~\cite{klosek-dygas_colored_1988} who, by means of a Fokker--Planck approach (and in a small-$\tau$ expansion), obtained many of the results derived later through a path-integral approach. Below we compare in detail our results to those obtained previously emphasizing technical differences, places where we go beyond the early results, and putting them in the context of active matter. As we show the focus on phenomena related to active matter allows us to uncover new phenomena which were not discussed in the early works.

To study the $D\to 0$ asymptotics we apply recent techniques from the theory of large deviations~\cite{FW2012}. These enable us to relate the escape problem and the steady-state distribution problem using standard results. In particular, if we consider a metastable state $x_{0}$ and its basin of attraction $\mathcal{B}_{0}$
both the mean escape time $\left\langle T_{\esc}\right\rangle $ and the invariant measure $\rho$ within $\mathcal{B}_{0}$ can be asymptotically computed from a single function 
$\Phi_{\tau}$ called the quasipotential as follows
\begin{equation*}
\begin{casesb}
\frac{\rho(x)}{\rho(x_{0})} & \underset{D\rightarrow0}{\asymp}e^{-\frac{\Phi_{\tau}(x)}{D}}\,,\\
\left\langle T_{\esc}\right\rangle  & \underset{D\rightarrow0}{\asymp}e^{\frac{\Phi_{\tau}(x_{\sa})}{D}}\,.
\end{casesb}
\end{equation*}
Here $x_{\sa}$ is the saddle point that needs to be crossed for the particle to escape from the basin of attraction
of $x_{0}$. The notation $f\underset{D\rightarrow0}{\asymp}g$
means logarithmic equivalence between the functions $f$ and $g$,
that is $\log f\underset{D\rightarrow0}{\sim}\log g$. Using standard tools we compute the quasipotential
$\Phi_{\tau}$ in both limits $\tau\rightarrow0$ (to order $\tau^2$) and $\tau\rightarrow+\infty$,
without any restriction on $U$. Our main findings are:
\begin{enumerate}

\item In the limit $\tau\rightarrow0$, we find the asymptotic expansion
\begin{equation}
\Phi_{\tau}(x)\underset{\tau\rightarrow 0}{\sim}U(x)-U(x_{0})+\tau\frac{\left|U'(x)\right|^{2}}{2}-\frac{\tau^{2}}{2}\int_{x_{0}}^{x}\left|U'(y)\right|^{2}U'''(y)\,{\rm d}y+O(\tau^{3})\,.\label{eq:quasipot small tau-1}
\end{equation}
The correction of order $\tau^{2}$ in $\Phi_{\tau}$ is the first
\textbf{nonlocal term} of the invariant measure. We interpret this
contribution in terms of long-range interactions in the system. These, as we show, lead to a
ratchet effect where particles are driven in the presence of an asymmetric potential (see Sec.~\ref{subsec:The-ratchet-effect} and Fig.~\ref{fig:ratchet})~\cite{Angelani2011}. We note that non-local interactions have been already observed in active fluids~\cite{baek2018generic} and therefore seem to be generic for this class of systems. The non-local contribution, interestingly, is absent for a harmonic potential. Our method can further be used to derive an
expansion for $\Phi_{\tau}$ to any order in $\tau$. We note from
a technical point of view that the expansion~(\ref{eq:quasipot small tau-1})
in powers of $\tau$ is non-intuitive: because of the existence of
a boundary layer in the instanton calculation used to evaluate $\Phi_{\tau}(x)$, a standard perturbative
procedure fails and the calculation has to be carried out with some care.
The singular behavior of the instanton also brings an insight on how an active optimal trajectory differs from an equilibrium one.

Comparing to earlier works, the $O(\tau)$ contribution to~(\ref{eq:quasipot small tau-1}) was derived by Wio \emph{et al}.~in~\cite{wio_path-integral_1989} using a path integral approach.
The escape rate implied by~(\ref{eq:quasipot small tau-1}) (\emph{i.e.}~for $x=x_{\sa}$) was derived in~\cite{bray_instanton_1989,bray_path_1990} at order $\tau^2$ and higher using path integrals. However, in the latter case the optimal path is not singular, and the perturbation is organized in even powers of $\tau$. The methods of~\cite{bray_instanton_1989,bray_path_1990} therefore allows one to determine the escape rate but not the steady-state quasipotential for any $x$.
However, as discussed at the end of Sec.~\ref{sec:AOUP-in-the}, we note that expression~(\ref{eq:quasipot small tau-1}) can be deduced from a result of Ref.~\cite{klosek-dygas_colored_1988} which use a Fokker--Planck approach.
This provides an alternative way to derive~(\ref{eq:quasipot small tau-1}), but one that does not uncover the singular nature of the instanton. The approach we follow gives an insight on the typical trajectory taken by the active particle to reach an arbitrary point.

\item The limit $\tau\rightarrow+\infty$ displays a different
behavior. The quasipotential should be seen as a functional of the
force field $F(x)=-U'(x)$, rather than a functional of the potential.
We show that the quasipotential has a singularity at the inflection
point $x_{\cri}$ where $U''(x_{\cri})=0$. Before the inflection point,
$\Phi_{\tau}$ is a local function of the force, and is given by 
\begin{equation}
\text{for }x<x_{\cri},\;\Phi_{\tau}(x)\underset{\tau\rightarrow+\infty}{\sim}\tau\frac{\left|U'(x)\right|^{2}}{2}+\frac{\left|U'(x)\right|^{2}}{2U''(x)}+O\left(1/\tau\right).\label{eq:expansion large tau 1-1}
\end{equation}
However, beyond the inflection point, it is entirely controlled by the vicinity
of $x_{\cri}$ 
\begin{equation}
\text{for }x>x_{\cri},\;\Phi_{\tau}(x)\underset{\tau\rightarrow+\infty}{\sim}\tau\frac{\left|U'(x_{\cri})\right|^{2}}{2}+\tau^{1/3}\frac{C\left|U'(x_{\cri})\right|^{2}}{\left(\sqrt{\frac{1}{2}\left|U'''(x_{\cri})\right|\left|U'(x_{\cri})\right|}\right)^{2/3}}+O(\tau^0)\,,\label{eq:expansion large tau 2-1}
\end{equation}
where $C\approx0.8120$ is a constant that can be computed numerically from the solution of a non-dimensional equation (see Eqs.~(\ref{eq:BL1-1}-\ref{eq:BL2-1}) and Sec.~\ref{sec:comments}).
This demonstrates again the non-locality of the problem.
The result implies that in the large $\tau$ limit, the potential barrier seen by the active
particle is controlled by forces. For example, in the escape problem the particle has
to overcome the maximal force $F_{{\rm max}}=U'(x_{\cri})$, as opposed to the equilibrium problem where maximal potential is the bottleneck. The proof
of Eq.~(\ref{eq:expansion large tau 2-1}) further reveals that the
dynamical system~(\ref{eq:AOUP model}) has a bifurcating slow manifold
that is used by the active particle to switch between the metastable
states. This particular kind of transition paths has been already
observed numerically for some multiscale systems in~\cite{grafke2017non}.
This gives rise to interesting new phenomena, which we discuss, such as hopping over
metastable states (see especially Sec.~\ref{subsec:Hopping-over-metastable} and the supplementary movies).

In comparison to previous works we note that the $\tau^{1/3}$ scaling of the correction~(\ref{eq:expansion large tau 2-1}) was hinted at in Ref.~\cite{bray_instanton_1989}. Expression~(\ref{eq:expansion large tau 2-1}), in the case $x=x_{\cri}$ useful in the escape problem, was derived in~\cite{bray_path_1990} through a path-integral approach, but again the singular nature of the instanton (that appears only for $x\neq x_{\cri}$) was not identified. Finally, we comment that the steady-state quasipotential cannot be inferred from the approach of Ref.~\cite{bray_path_1990}.

\end{enumerate}

The paper is organized as follows: for completeness we first give in Section~\ref{sec:Summary-of-the} a summary of the approach used for the calculations, before discussing their consequences. 
The detailed derivations of the results are given in Sections~\ref{sec:AOUP-in-the} and~\ref{sec:AOUP-in-the-large}.

\section{The approach and some implications of the results\label{sec:Summary-of-the}}

\subsection{Summary of the large deviation framework\label{subsec:Summary-of-the}}

In this section we describe the approach used to evaluate the quasipotential $\Phi_{\tau}$. For completeness, the relation between the mean escape time and the quasipotential is recalled in
Appendix~\ref{sec:Mean-escape-time}. In what follows, we assume for simplicity that $\rho(x_{0})=1$, the generalization being straightforward. First, note that the invariant measure can be expressed in terms of the transition probability of the AOUP as
\begin{equation*}
\rho(x)=\underset{T\rightarrow-\infty}{\lim}\int_{-\infty}^{+\infty} P(x,v,0|x_{0},0,-T)\;{\rm d}v \,,
\end{equation*}
where $P(x,v,0|x_{0},0,-T)$ is the probability to be at $(x,v)$
at time $t=0$ starting initially at the stationary point $(x_{0},0)$
at $t=-T$. In this paper, we use both the Lagrangian and the Hamiltonian formalism, and we briefly review both in the following.
\paragraph{Lagrangian formalism: }
The transition probability can then be expressed as an Onsager--Machlup
path integral~\cite{onsager_fluctuations_1953}. This gives for the invariant measure
\begin{equation}
\rho(x)=\int\mathcal{D}\left[x(t),v(t)\right]e^{-\frac{1}{D}\mathcal{A}[x,v]},\label{eq:path integral}
\end{equation}
with
\begin{align}
\mathcal{A}[x,v]&=\int_{-\infty}^{0}\mathcal{L}\left(x,v,\dot{x},\dot{v}\right)\dd t\label{eq:action}\\
\mathcal{L}\left(x,v,\dot{x},\dot{v}\right)&=\begin{cases}
\frac{1}{4}\left(\tau\dot{v}+v\right)^{2} & \text{if }v=\dot{x}+U'(x)\\
+\infty & \text{otherwise} \;.
\end{cases}\label{eq:Lagrangien}
\end{align}
Using the constraint on $v$ the Lagrangian can be equivalently written as
\begin{equation}
\mathcal L(x,\dot x,\ddot{x}) = 
\frac{1}{4} \Big(\dot{x}+U'(x) + \tau  \big(\ddot{x}+\dot{x} U''(x)\big) \Big)^2 .
\label{eq:LagLag2}
\end{equation}
The instanton trajectory is found by solving the Euler--Lagrange equation with the boundary conditions
\begin{align}
x(t)\underset{t\rightarrow-\infty}{\longrightarrow}x_{0}; & ~\; v(t)\underset{t\rightarrow-\infty}{\longrightarrow}0\nonumber  \\
x(0)=x; & ~\; \dot{x}(0)=0
\,.
\label{eq:BC}
\end{align}
The first line implies that the particle starts at the stationary state, while the second assures that it arrives at $x$ at time zero without overshooting. So far, expression~(\ref{eq:path integral}) is formally exact (with a proper interpretation/discretization of the action, see \emph{e.g.}~\cite{cugliandolo_building_2019}). In
the limit $D\rightarrow0$, one can use a saddle-point approximation to get
\begin{equation*}
\int\mathcal{D}\left[x(t),v(t)\right]e^{-\frac{1}{D}\mathcal{A}[x,v]}\underset{D\rightarrow 0}{\asymp}e^{-\frac{1}{D}\Phi_{\tau}(x)},
\end{equation*}
where $\Phi_{\tau}(x)$ is the quasipotential.

\paragraph{Hamiltonian formalism: } The easiest path for obtaining the Hamiltonian formalism is through the Martin--Siggia--Rose--Janssen--De~Dominicis approach~\cite{Janssen1976,Janssen1979,dominicis_techniques_1976,DeDominicis1978,Martin1973}. To do so we rewrite the probability by reexpressing delta functions as 
\begin{equation}
	\rho(x)= \bigg\langle \int \mathcal{D}\left[x(t),v(t),P_v(t),P_x(t)\right]  
        e^{-\frac{1}{D}\int_{-\infty}^0  \big\lbrace P_x(\dot{x}-v+U')+P_v(\tau \dot{v}-v-\sqrt{2D}\xi(t)) \big\rbrace \dd t }\bigg\rangle_\xi
        \;,
\end{equation}
which after averaging over the noise $\xi$ gives
\begin{equation}
	\rho(x)=  \int \mathcal{D}\left[x(t),v(t),P_v(t),P_x(t)\right]  e^{-\frac{1}{D}\int_{-\infty}^0 \big\lbrace P_x \dot{x}+P_v \dot{v}-H[P_x,P_v,x,v] \big \rbrace \dd t }\;, \label{eq:Hpath}
\end{equation}
where the Hamiltonian is given by
\begin{equation}
H\left(x,v,P_{x},P_{v}\right)=\left(v-U'(x)\right)P_{x}-\frac{vP_{v}}{\tau}+\frac{P_{v}^{2}}{\tau^{2}} \,.\label{eq:hamiltonian}
\end{equation}
The instanton is then given by a solution of Hamilton's equations
\begin{equation}
\begin{casesb}
\dot{x} & = v-U'(x)\\
\dot{v} & = -\frac{v}{\tau}+2\frac{P_{v}}{\tau}
\end{casesb}
\qquad
\text{and}
\qquad
\begin{casesb}
\dot{P_{x}} & = U''(x)P_{x}\\
\dot{P_{v}} & = \frac{P_{v}}{\tau}-P_{x} 
\end{casesb}\label{eq:hamilton space}
\ ,
\end{equation}
with the boundary conditions 
\begin{align}
x(t)\underset{t\rightarrow-\infty}{\longrightarrow}x_{0};  ~\;& v(t),P_{x}(t),P_{v}(t)\underset{t\rightarrow-\infty}{\longrightarrow}0\,. \label{eq:BC1} \\
P_{v}(0)&=0\label{eq:BCPv}
\end{align}
The boundary condition on $P_v(0)$ can be deduced by noting that $P_v=\partial_v S$, with $S$ the action corresponding to $H$, and that we are minimizing the action with respect to the final velocity. 
It is equivalent to the Lagrangian one, Eq.~(\ref{eq:BC}).
The boundary conditions at $t = -\infty$ results from demanding that we start at the stationary point.

Finally, note that the Hamiltonian has no explicit time dependence, `energy' is conserved
along the optimal path. Evaluating the Hamiltonian~(\ref{eq:hamiltonian})
at $t=-\infty$ gives the constraint
\begin{equation}
H\left(x,v,P_{x},P_{v}\right)=0
\,.\label{eq:Hconservation}
\end{equation}
With the above we now turn to discuss the small and large $\tau$ expansions.

\subsection{The small $\tau$ limit\label{subsec:The-small-}}

We now explain how the value of the quasipotential at
the saddle point $\Phi_{\tau}(x_{\sa})$ can, for the escape problem, be evaluated to order
$\tau^{2}$ using a standard perturbation theory~\cite{bray_instanton_1989,bray_path_1990}. 
Note that this approach, as we discuss in Sec.~\ref{sec:expansionorderbyorder}, 
in fact fails when the final point $x$ is not a saddle point $x_{\sa}$, \emph{i.e.}~for the computation of the invariant measure  $\Phi_{\tau}(x)$ at an arbitrary point $x$. We will show
how the perturbation scheme has to be modified in this case. For clarity, these subtleties will be illustrated for a harmonic potential
$U(x)=\frac{1}{2}kx^{2}$ in paragraph~\ref{sec:harmonic}, where $\Phi_{\tau}(x)$ is computed exactly. In what follows we always take $U(x_{0})=0$.

To proceed, we first expand the square in the action~(\ref{eq:action}-\ref{eq:Lagrangien}) so as to write
\begin{equation*}
\mathcal{A}[x]=\mathcal{A}_{0}[x]+\tau\mathcal{A}_{1}[x]+\tau^{2}\mathcal{A}_{2}[x] \,. 
\end{equation*}
Note that Eq.~(\ref{eq:Lagrangien}) implies that $\mathcal{A}_{1}[x]=\frac{1}{2}\int_{-\infty}^{0}\dot{v}v \,{\rm d}t=\frac{1}{4}v(0)^{2}$
is a boundary term. Furthermore, since we are looking for action minimizers which reach
the saddle $x_{\sa}$, the fluctuation path $\tilde{x}$ (or optimal path)
has to reach $x_{\sa}$ with zero velocity. This can be directly seen from Eq.~(\ref{eq:Hconservation}) with the boundary condition Eq.~(\ref{eq:BCPv}). Therefore, for $x_{\sa}$ a saddle, $\mathcal{A}_{1}[\tilde{x}]=0$. Next, using Eq.~(\ref{eq:Lagrangien})
\begin{equation}
\begin{casesb}
\mathcal{A}_{0}[x] & = \frac{1}{4}\int_{-\infty}^{0}v^{2}\,{\rm d}t \,,\\
\mathcal{A}_{2}[x] & = \frac{1}{4}\int_{-\infty}^{0}\dot{v}^{2}\,{\rm d}t \,.
\end{casesb}\label{eq:power action}
\end{equation}
As the action expansion has no terms proportional to $\tau$, we can
expand the fluctuation path $\tilde{x}(t)$ and the quasipotential
$\Phi_{\tau}(x_{\sa})$ in powers of $\tau^{2}$
\begin{align}
\tilde{x}(t) & =\tilde{x}_{0}(t)+\tau^{2}\tilde{x}_{2}(t)+O(\tau^{4})\\
\Phi_{\tau}(x_{\sa}) & =\Phi_{0}(x_{\sa})+\tau^{2}\Phi_{2}(x_{\sa})+O(\tau^{4})
\label{eq:expansionPhitau}
\end{align}
and solve the minimization problem up to order $\tau^{2}$. The
zeroth order is given by the equilibrium fluctuation path
\begin{equation}
\begin{casesb}
\dot{\tilde{x}}_{0}(t) & = U'(\tilde{x}_{0}(t))\,,\\
\mathcal{A}_{0}[\tilde{x}_{0}] & = U(x_{\sa})\,,
\end{casesb}\label{eq:0fluctuation}
\end{equation}
and the mean escape time is given by the classical Arrhenius law,
with the quasipotential
\begin{equation*}
\Phi_{0}(x_{\sa})=U(x_{\sa}) - U(x_0)
\,.
\end{equation*}
The second order correction to the quasipotential is given by
\begin{equation*}
\Phi_{2}(x_{\sa})=\int\frac{\delta\mathcal{A}_{0}}{\delta x}[\tilde{x}_{0}(t)]\,\tilde{x}_{2}(t)\,{\rm d}t
\ + \ 
\mathcal{A}_{2}[\tilde{x}_{0}(t)]\,.
\end{equation*}
Because $\tilde{x}_{0}(t)$ is a minimizer of $\mathcal{A}_{0}$,
the first term vanishes and we are left with
\begin{align}
\Phi_{2}(x_{\sa}) & =\mathcal{A}_{2}[\tilde{x}_{0}(t)]
=\int_{-\infty}^{0}\left(U'U''\right)^{2}{\rm d}t \,,
\label{eq:Phi2timedep}
\end{align}
where the integral has to be evaluated along the equilibrium fluctuation path~(\ref{eq:0fluctuation}).
Finally, using ${\rm d}t=\frac{{\rm d}x}{|U'|}$, see Eq.~(\ref{eq:0fluctuation}), we obtain
\begin{equation}
\Phi_{2}(x_{\sa})=\int_{x_{0}}^{x_{\sa}}\left|U'(y)\right|\left(U''(y)\right)^{2}{\rm d}y \,. \label{eq:2quasipot}
\end{equation}
%
%
This relation gives the correction to order $\tau^{2}$
for the Arrhenius mean escape time, as $\left\langle T_{\esc}\right\rangle \underset{D\rightarrow0}{\asymp}e^{\frac{\Phi_{\tau}(x_{\sa})}{D}}$ with the expansion given in Eq.~(\ref{eq:expansionPhitau}).

As mentioned above the perturbative procedure described above  and the result
of Eq.~(\ref{eq:2quasipot}) (known since the works of Bray, MacKane and collaborators~\cite{bray_instanton_1989,bray_path_1990}) \textbf{cannot} be used, in a straightforward manner, to compute the perturbative corrections to the invariant measure. The reason is that
the fluctuation path which reaches a point~$x$ that is not a saddle (\emph{i.e.}~$x\neq x_{\sa}$) displays a
boundary layer of size $\tau$ close to $t=0$. This peculiarity
of the fluctuation path gives local contributions to the quasipotential
to any orders in $\tau$, that only vanish at $x=x_{\sa}$. This effect is already present for the exactly solvable case of harmonic potential as illustrated in Sec.~\ref{sec:harmonic}. The full
computation of the fluctuation path and the quasipotential for a general potential up to order
$\tau^{2}$ is rather lengthy, and is detailed in Sec.~\ref{sec:AOUP-in-the}. 
The final result was announced in Eq.~(\ref{eq:quasipot small tau-1}).
As emphasized previously, Eq.~(\ref{eq:quasipot small tau-1}) shows that
the second order correction to the quasipotential is \emph{non-local}. Several hallmarks characterizing active systems, such as ratchet currents, originate from such
contributions. We illustrate this on the ratchet example in Sec.~\ref{subsec:The-ratchet-effect}.

\subsubsection{Instanton path for the harmonic potential}
\label{sec:harmonic}

To illustrate the subtle boundary layer that emerges in the instanton path for $\tau>0$, we consider the AOUP in a harmonic potential $U(x)=\frac{1}{2}kx^{2}$.
In this case, both the fluctuation path and the quasipotential can
be computed exactly. Interestingly, all terms to order larger than
$\tau$ in the expansion vanish. Therefore, the problem reduces to an effective equilibrium Brownian
particle with an ``effective temperature'' that depends on $\tau$ and the stiffness of the trap $k$ (see \emph{e.g.}~\cite{woillez2019active}).

\begin{figure}
\begin{centering}
\includegraphics[width=9cm]{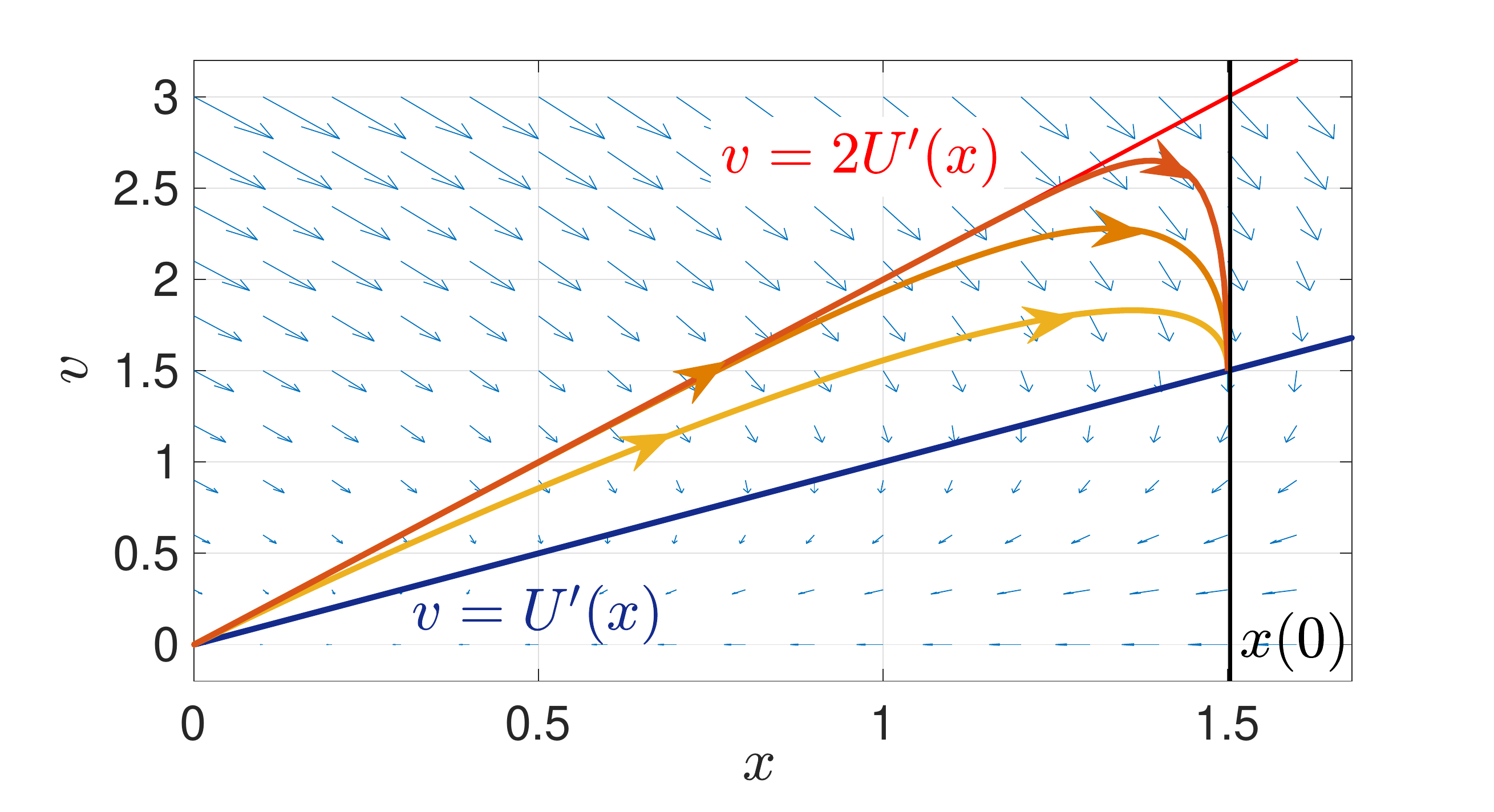}\includegraphics[width=9cm]{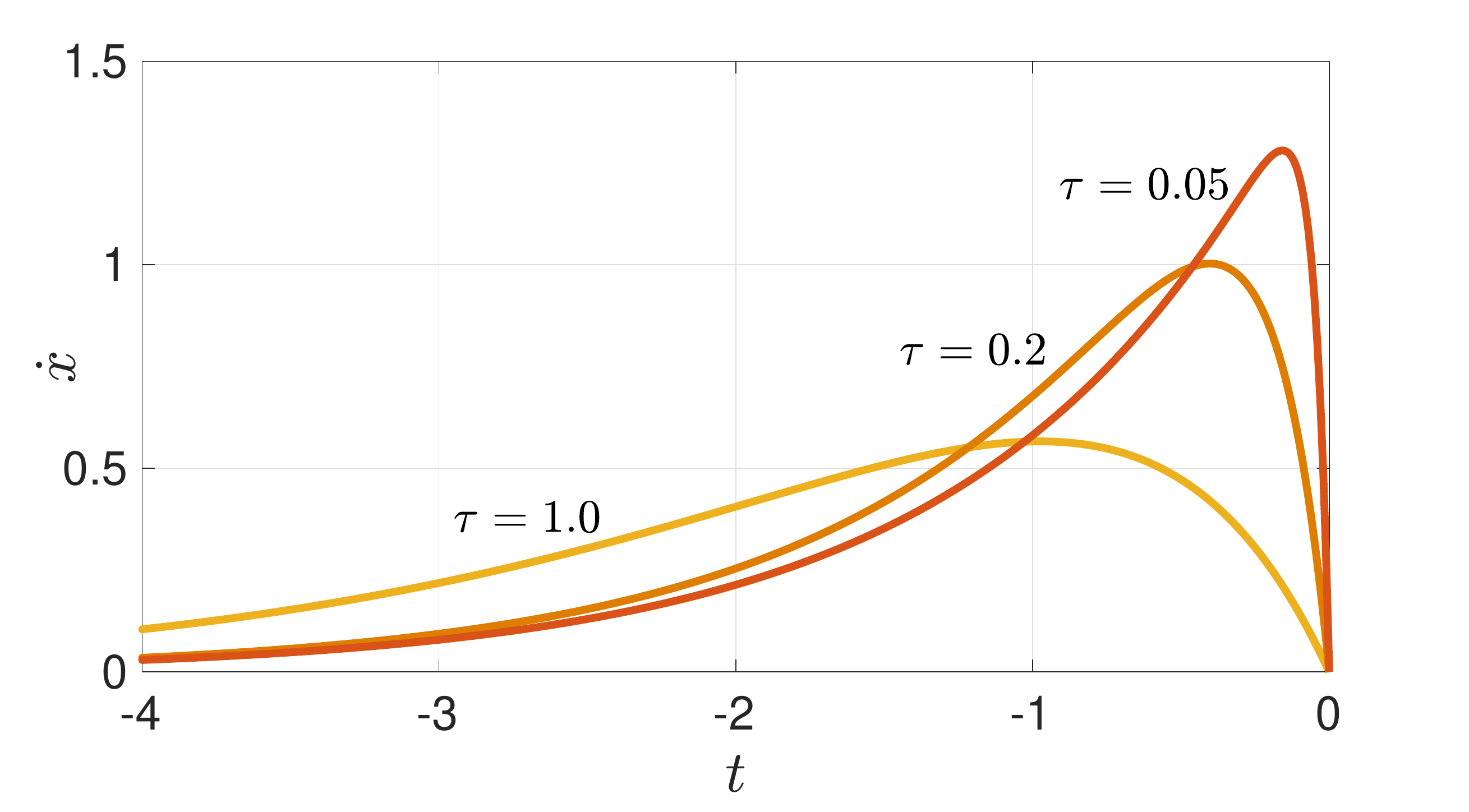}
\par\end{centering}
\caption{Fluctuation path for the AOUP in a harmonic potential $U(x)=\frac{x^{2}}{2}$.
\textbf{Left: }The graph displays the fluctuation paths in the phase
space $(x,v)$, for the three values $\tau=1.0/0.2/0.05$. The light
blue arrows represent the vector field for $\tau=0.5$. The dark blue
line gives the conditions for spatial stationary state $\dot{x}=0$,
and the red line is the equation $v=2U'(x)$ for the equilibrium fluctuation
path (\emph{i.e.}~$\tau=0$). It can be seen that the convergence to the equilibrium path
in the limit $\tau\rightarrow0$ displays a singularity close to $x(0)=1.5$.
\textbf{Right:} The trajectories $\dot{x}(t)$ for the same values
of $\tau$ as in the left picture. The fluctuation path satisfies $\dot{x}(0)=0$,
whatever the value of $\tau$. The boundary layer of size $\propto\tau$
can clearly be identified close to $t=0$. \label{fig:harmonic fluctuations}}

\end{figure}

To see this, define the a variable $p=v-kx$, so that Eqs.~(\ref{eq:AOUP model}) become
\begin{equation*}
\begin{casesb}
\dot{x} & = p,\\
\dot{p} & = -\left(\frac{1}{\tau}+k\right)p-\frac{k}{\tau}x+\sqrt{\frac{2D}{\tau^{2}}}\xi(t)\,.
\end{casesb}
\end{equation*}
Then, we can identify a friction coefficient $\gamma=\frac{1}{\tau}+k$,
and effective potential $U_{{\rm eff}}(x)=\frac{kx^{2}}{2\tau}$ so that
\begin{equation}
\begin{casesb}
\dot{x} & = p,\\
\dot{p} & = -\gamma p-\nabla U_{{\rm eff}}+\sqrt{\frac{2\gamma D}{\tau+k\tau^{2}}}\xi(t)\,.
\end{casesb}\label{eq:brownian}
\end{equation}
Eqs.~(\ref{eq:brownian}) are exactly the equations of an underdamped equilibrium
Brownian particle in a potential $U_{{\rm eff}}(x)$ and a
temperature $T_{{\rm eff}}=\frac{D}{\tau+k\tau^{2}}$. The problem
can therefore be solved exactly:
\begin{enumerate}
\item The \textbf{mean escape time} satisfies the Arrhenius law
\begin{equation}
\left\langle T_{\esc}\right\rangle \asymp e^{\frac{U_{{\rm eff}}}{T_{{\rm eff}}}}=e^{\frac{\left(1+\tau k\right)}{D}k\frac{x^{2}}{2}}\,,\label{eq:Tesc harmonic}
\end{equation}
which corresponds to Eq.~(\ref{eq:quasipot small tau-1}) where all terms of order $\tau^2$ are zero.
\item The \textbf{fluctuation path} is obtained using the time reversal symmetry of the problem, namely, by reversing the friction term in the relaxation (noiseless) equations
\begin{equation*}
\begin{casesb}
\dot{x} & = p\,,\\
\dot{p} & = +\gamma p-\nabla U_{{\rm eff}}\,,
\end{casesb}
\end{equation*}
which equivalently gives 
\begin{equation}
\ddot{x}-\Big(k+\frac{1}{\tau}\Big)\dot{x}+\frac{k}{\tau}x=0\,.\label{eq:fluc path harmonic}
\end{equation}
\end{enumerate}
Eq.~(\ref{eq:fluc path harmonic}) together with the boundary conditions
$x(0)=x$ and $\dot{x}(0)=0$ gives
\begin{equation}
x(t)=\frac{x}{1-k\tau}\left(e^{kt}-k\tau e^{\frac{t}{\tau}}\right).\label{eq:solution harmonic}
\end{equation}
The `bulk' contribution (in the regime $|t|\gg \tau$) and the boundary layer contribution (in the regime $|t|\sim \tau$) are easily
identified in Eq.~(\ref{eq:solution harmonic}) as the two exponential
terms. Figure~\ref{fig:harmonic fluctuations} displays the fluctuation
path~(\ref{eq:solution harmonic}) for decreasing values of $\tau$.
It is clearly seen that the path comes closer to the equilibrium
fluctuation path of equation $\dot{x}=kx$, when $\tau$ decreases towards $0$.
However, the fluctuation path always has to satisfy the boundary condition
$\dot{x}(0)=0$: this constraint is responsible for the existence of
the boundary layer of a size of order $\tau$, and creates a singularity
at $t=0$ in the fluctuation path in the limit $\tau\rightarrow0$.
This surprising phenomenon explains the breakdown of the standard
perturbative expansion of the fluctuation path, and creates, as we show, non-trivial
local contributions to the quasipotential Eq.~(\ref{eq:quasipot small tau-1})
to all orders in $\tau$.

\subsubsection{A ratchet\label{subsec:The-ratchet-effect}}

\begin{figure}
\begin{centering}
\includegraphics[width=15cm]{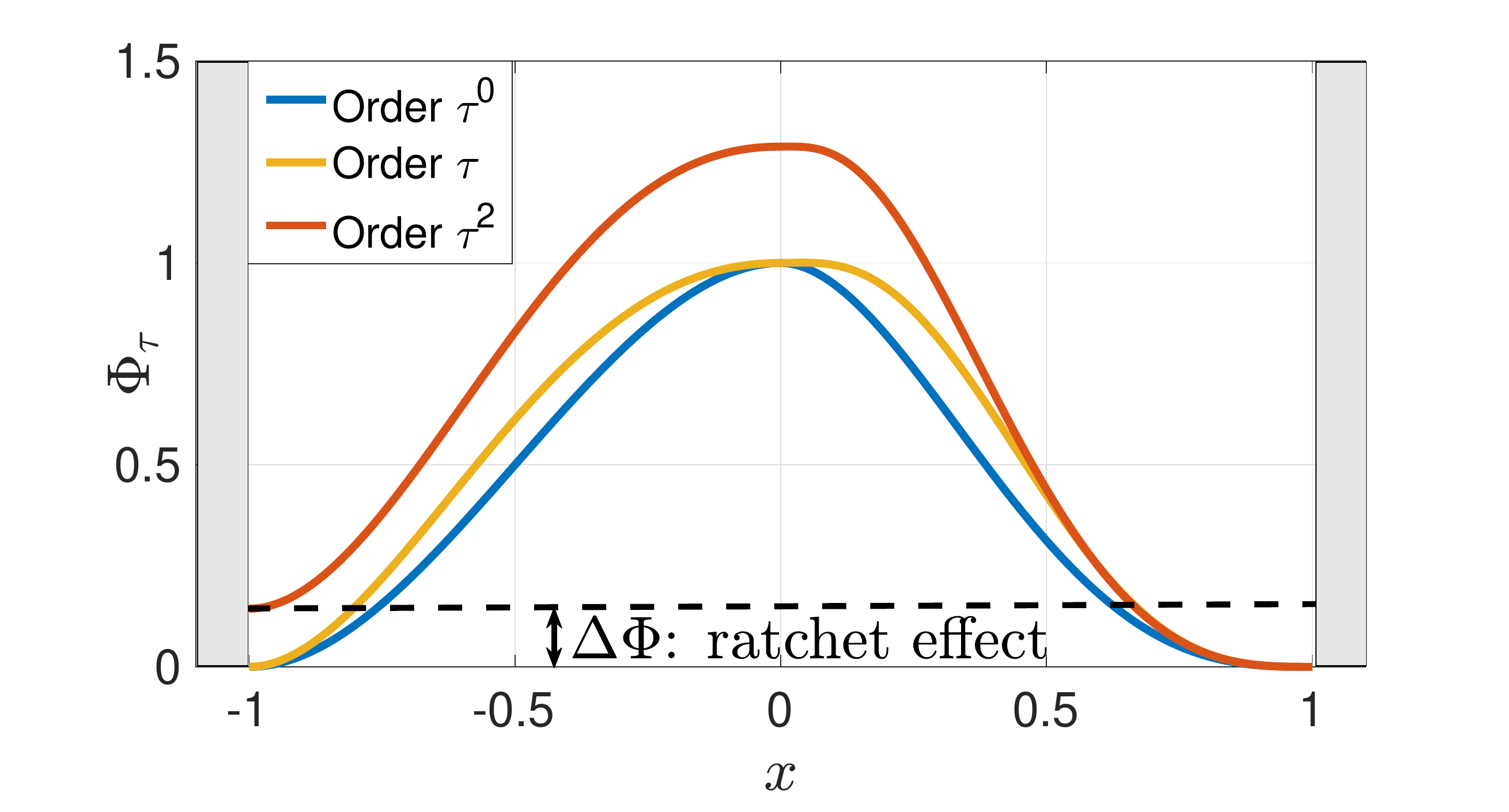}
\par\end{centering}
\caption{The ratchet effect for AOUP. The blue curve is the ratchet potential
$U(x)$, which also corresponds to the quasipotential at equilibrium
$\Phi_{\tau=0}$. The figure displays the quasipotential $\Phi_{\tau}$
truncated to order $\tau$ (yellow curve), and $\tau^{2}$ (red curve),
for $\tau=0.1$. It can be seen that an offset $\Delta\Phi$ appears
between the right and the left of the ratchet to order $\tau^{2}$,
creating an accumulation of active particles on the right of the ratchet.
\label{fig:ratchet}}

\end{figure}

We next illustrate that a consequence of the non-local contribution to the quasipotential
(\ref{eq:quasipot small tau-1}) is a ratchet effect. We consider the
AOUP in the domain $[-1,1]$ with reflecting boundary conditions (walls)
at $x=\pm1$. We choose the continuous asymmetric potential 
\begin{equation*}
U(x)=\begin{cases}
1-3x^{2}-2x^{3} & \text{for }x\in[-1,0]\,,\\
1-6x^{2}+8x^{3}-3x^{4} & \text{for }x\in[0,1]\,.
\end{cases}
\end{equation*}
The function $U(x)$ is the blue curve displayed in Fig.~\ref{fig:ratchet}.
For this problem, we compute, using Eq.~(\ref{eq:quasipot small tau-1}),
the quasipotential $\Phi_{\tau}$ defined by the relation
\begin{equation*}
D\log\rho(x)=\Phi_{\tau}(x)+C\,,
\end{equation*}
where $C$ is some additional constant. Figure~\ref{fig:ratchet}
displays the function $\Phi_{\tau}(x)$ truncated respectively to
orders $\tau^{0}$, $\tau$ and $\tau^{2}$. It is shown that to order
$\tau^{2}$, there is an offset $\Delta\Phi$ given by 
\begin{align*}
\Delta\Phi & =\Phi_{\tau}(-1)-\Phi_{\tau}(1)\\
 & =-\tau^{2}\int_{-1}^{1}\left|U'(y)\right|^{2}U'''(y)\,{\rm d}y\,.
\end{align*}
The offset is responsible for the accumulation of active particles
on the right side of the ratchet potential, according to the relation
$\frac{\rho(1)}{\rho(-1)}\underset{D\rightarrow0}{\asymp}e^{\frac{\Delta\Phi}{D}}$.
This effect is absent if one considers the expansion of
$\Phi_{\tau}$ up to order $\tau$ only.

\subsection{The large $\tau$ limit\label{subsec:The-large-}}

\begin{figure}
\begin{centering}
\includegraphics[width=15cm]{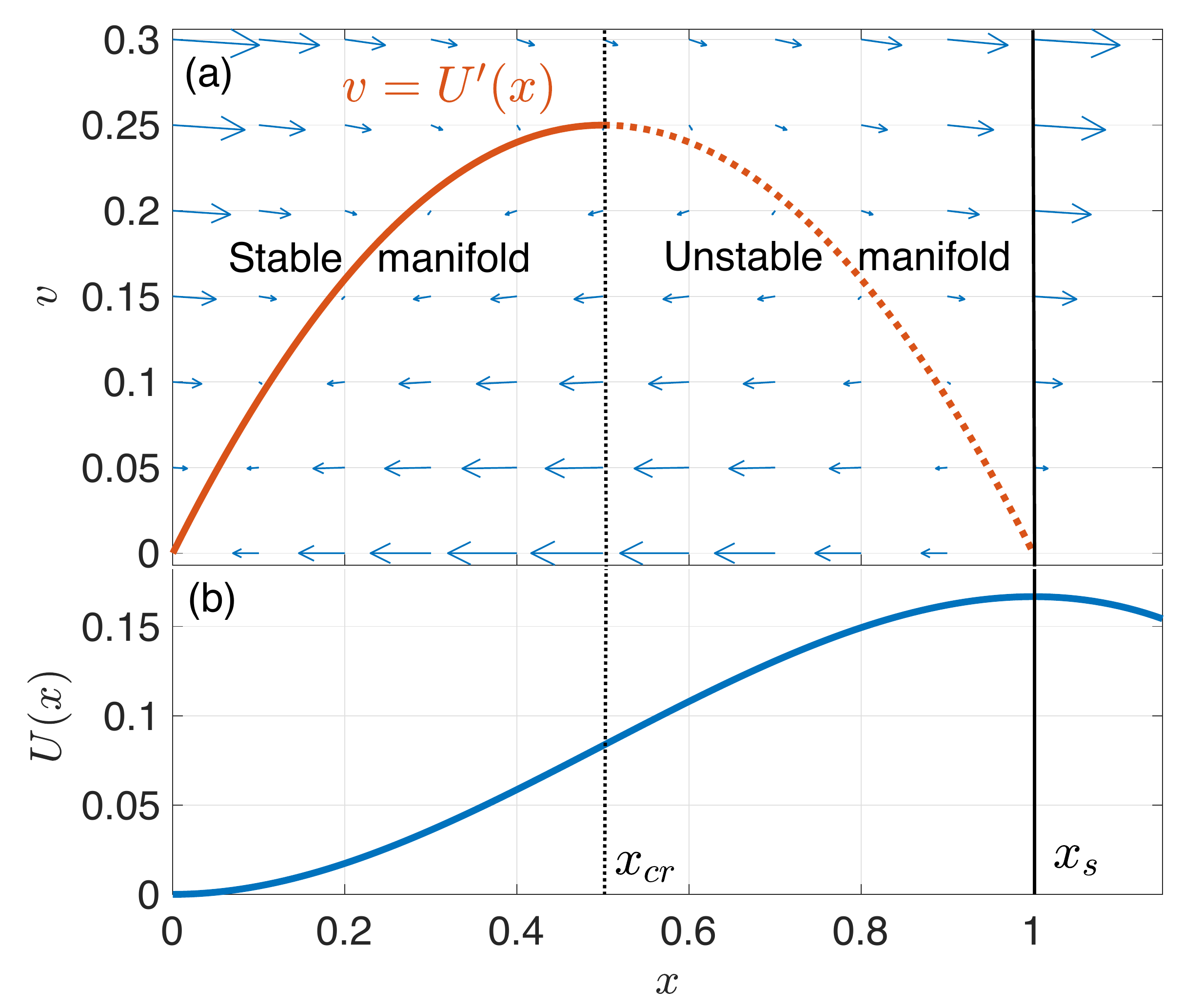}
\par\end{centering}
\caption{(a): Phase space structure of AOUP in the large $\tau$ limit. The dynamics
along $x$ is very fast compared to the $v$ direction. The blue arrows
represent the deterministic dynamics for $\tau=30$, and $U(x)=\frac{x^{2}}{2}-\frac{x^{3}}{3}$.
To escape from the metastable state, the process first follows the
stable manifold up to $x_{\cri}$ and then can fly over the barrier
using the deterministic field. (b): Graph of the potential barrier $U(x)$. The saddle point is $x_\sa=1$ and the inflection point is $x_{\cri}=0.5$. \label{fig:large tau barrier}}
\end{figure}

We now turn to discuss the large $\tau$ limit. In particular, we illustrate how results can be obtained using heuristic arguments. The full details of the calculation are presented in Sec.~\ref{sec:AOUP-in-the-large}. The limit is best understood by rescaling time according to $t\leftarrow t/\tau$
and $D\leftarrow D/\tau$ in Eqs.~(\ref{eq:AOUP model}). This yields
\begin{equation}
\begin{casesb}
\dot{x} & = \tau\left(v-U'(x)\right),\\
\dot{v} & = -v+\sqrt{2D}\xi(t)\,.
\end{casesb}\label{eq:AOUP large tau}
\end{equation}
From this it is easy to see that in the large $\tau$ limit the dynamical relaxation time
in the trap is much smaller than the time scale associated with the variation of the active
force. In particular, in the strict limit $\tau \to \infty$ the particle is in a quasistationary state given
by the equation
\begin{equation}
U'(x(t))=v(t)\,.\label{eq:quasistationary state}
\end{equation}
Since the dynamics of $v$ does not depend on $x$, the location of the
particle can be determined from Eq.~(\ref{eq:quasistationary state}) by
inverting the function $U'$. Figure~\ref{fig:large tau barrier}
displays the dynamics in phase space $(x,v)$ for $\tau=30$. This implies that, for a typical barrier, the curve $v=U'(x)$
has a stable manifold in an interval $[x_{0},x_{\cri}]$ and an unstable
manifold in $[x_{\cri},x_{\sa}]$. Here $x_{\cri}$ is the (unique) inflection
point at which $U''(x_{\cri})=0$. 

This implies that in order to cross the barrier, the particle has
to reach $x_{\cri}$ by moving along the stable manifold, and then, once $x_{\cri}$ is reached, it can
``fly over'' the barrier using the deterministic dynamics. Namely, any position
$x>x_{\cri}$ can be reached without any additional action cost, once
the inflection point $x_{\cri}$ has been crossed. We note that this
phenomenology seems to be generic for fluctuation paths in the presence
of a slow manifold as shown in~\cite{grafke2017non}.

The barrier displayed in Fig.~\ref{fig:large tau barrier} is
thus a \textbf{force barrier} and not an energy barrier as found in equilibrium problems. Then the mean escape time in the large deviation
regime $D\rightarrow0$ is given by the mean time required for the
Ornstein--Uhlenbeck process $v(t)$ to reach $F_{{\rm max}}=-U'(x_{\cri})$. The result is
\begin{equation}
\left\langle T_{\esc}\right\rangle \underset{D\rightarrow0}{\asymp}e^{\frac{\left|F_{{\rm max}}\right|^{2}}{D}}\,.\label{eq:escape result  large tau}
\end{equation}
Following the same line of arguments, one can obtain the quasipotential
in the large $\tau$ limit
\begin{equation}
\Phi_{\tau}(x)\underset{\tau\rightarrow+\infty}{\longrightarrow}\begin{cases}
\left|U'(x)\right|^{2} & \text{for }x_{0}<x<x_{\cri}\,,\\
\left|U'(x_{\cri})\right|^{2} & \text{for }x_{\cri}<x<x_{\sa}\,.
\end{cases}\label{eq:quasipot large tau}
\end{equation}

Whereas the result of Eq.~(\ref{eq:quasipot large tau}) is relatively
easy to obtain, the first order correction is not. One has to deal
with two different types of boundary layers in the fluctuation path,
depending on the value of $x$.
\begin{itemize}
\item For $x<x_{\cri}$, the fluctuation path has a boundary layer of size
$\frac{1}{\tau}$ close to $t=0$. The expansion of $\Phi_{\tau}$
can be done in powers of $\frac{1}{\tau}$, and one obtains to leading
order
\begin{equation}
\Phi_{\tau}(x)\underset{\tau\rightarrow+\infty}{\sim}\frac{\left|U'(x)\right|^{2}}{2}+\frac{\left|U'(x)\right|^{2}}{2\tau U''(x)}+O\left(1/\tau^{2}\right).\label{eq:expansion large tau 1}
\end{equation}
Note that as required this expression agrees with the exact expression~(\ref{eq:Tesc harmonic})
obtained  for a harmonic potential.
\item For $x>x_{\cri}$, the boundary layer is located in the vicinity of
$x_{\cri}$, with the particular scaling $\frac{1}{\tau^{2/3}}$: this implies
in particular  that the correction depends only on the expansion
of $U$ close to $x_{\cri}$. We obtain
\begin{equation}
\Phi_{\tau}(x)\underset{\tau\rightarrow+\infty}{\sim}\frac{\left|U'(x_{\cri})\right|^{2}}{2}+\frac{C\left|U'(x_{\cri})\right|^{2}}{\left(\tau\sqrt{\frac{1}{2}\left|U'''(x_{\cri})\right|\left|U'(x_{\cri})\right|}\right)^{2/3}}+O(1/\tau)\,,\label{eq:expansion large tau 2}
\end{equation}
where $C\approx0.8120$ is a constant that can be 
computed numerically from the solution of a non-dimensional equation (see Eqs.~(\ref{eq:BL1-1}-\ref{eq:BL2-1})  and paragraph~\ref{sec:comments}).
This result was derived in~\cite{bray_path_1990} for the special case of $x=x_{\cri}$ that one has to consider for the escape problem. The singular nature of the instanton is absent in this special situation.
\end{itemize}

\subsubsection{Hopping over metastable states\label{subsec:Hopping-over-metastable}}

\begin{figure}
\includegraphics[width=9cm]{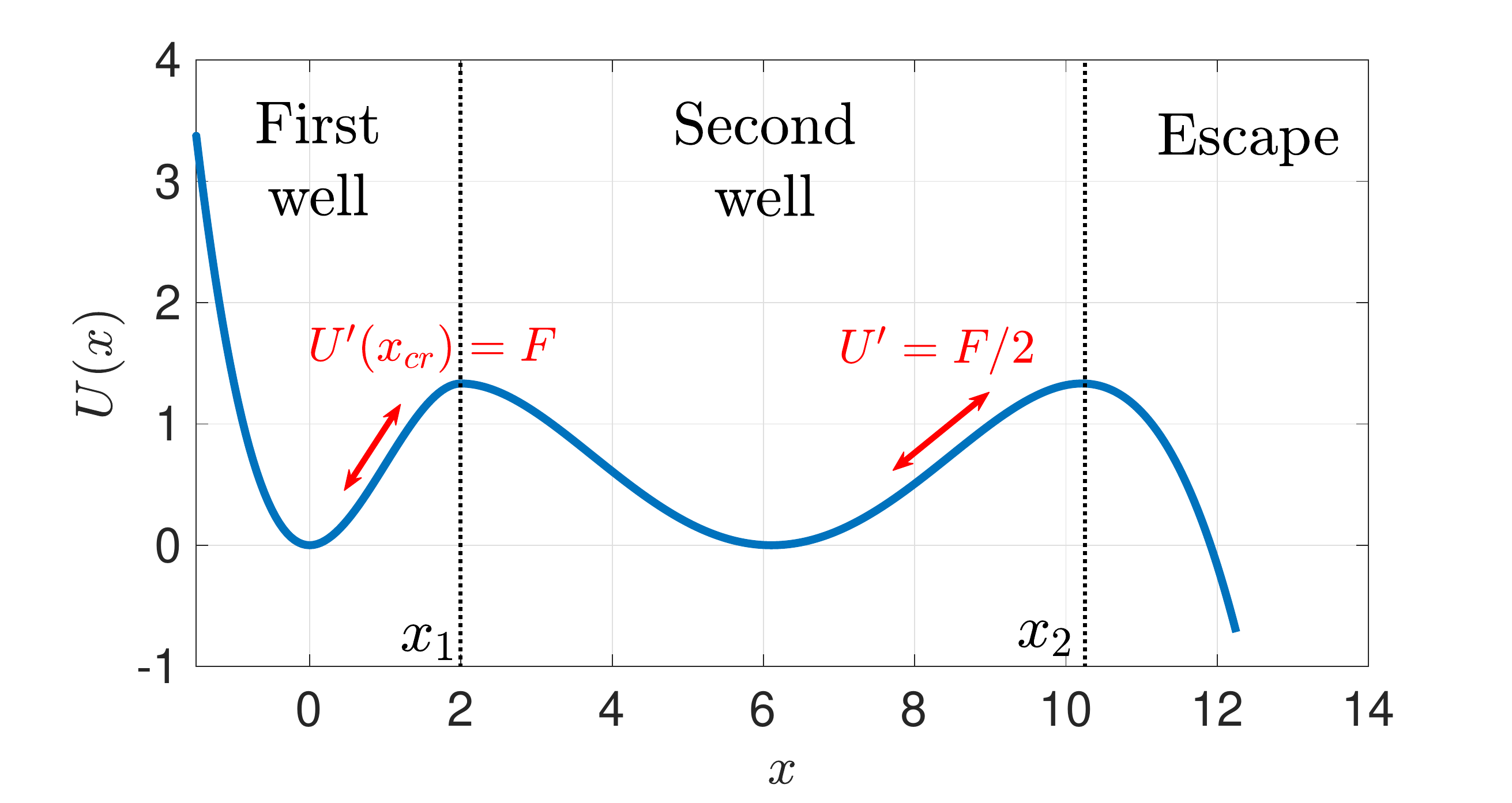}\includegraphics[width=9cm]{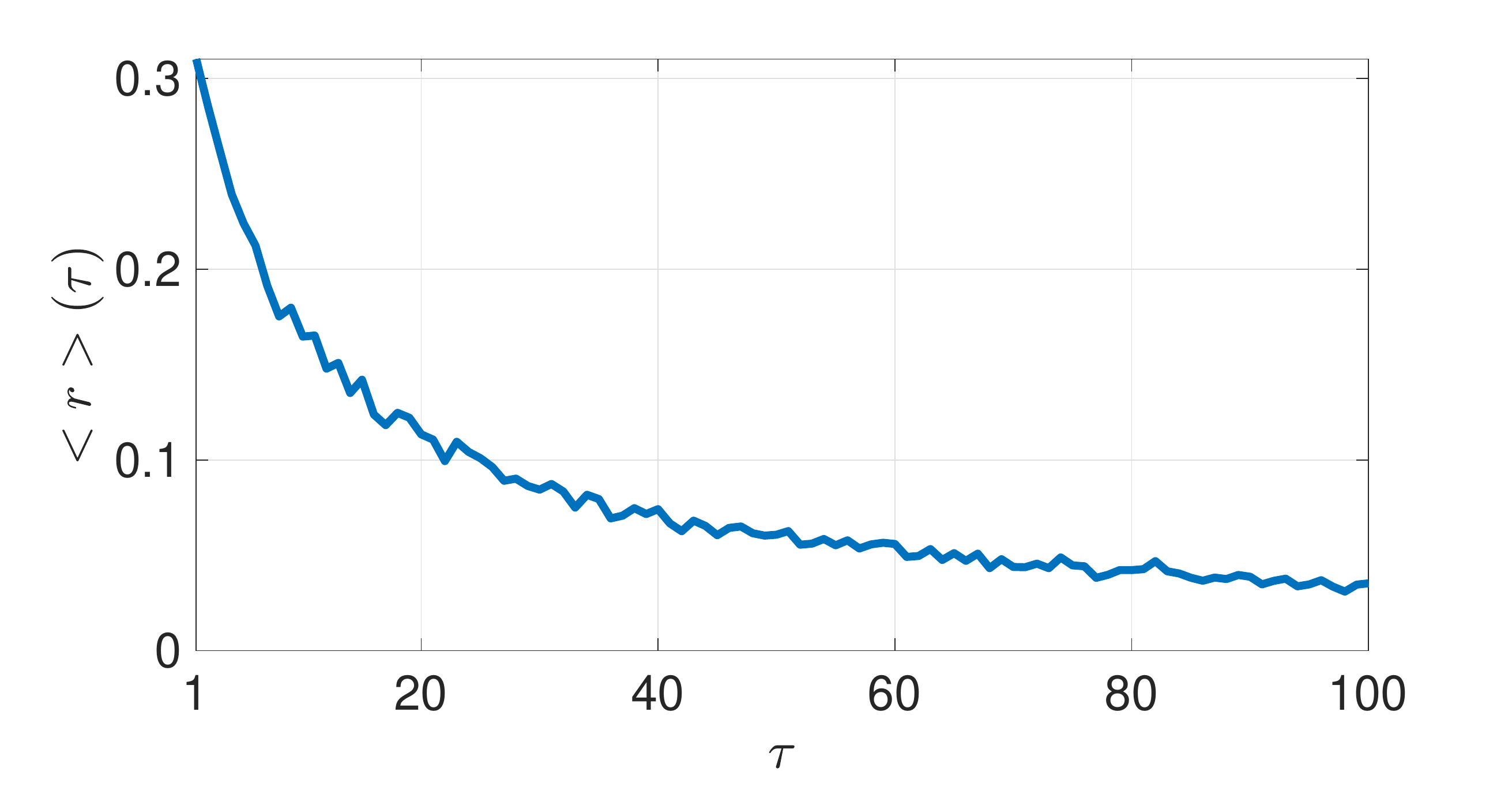}

\caption{\textbf{Left}: The metastable potential $U(x)$. The AOUP particle
has to cross two barriers at $x_{1}$ and $x_{2}$ to escape. The
two wells are designed such that the maximal slope of the first barrier
is $F$, whereas the maximal slope of the second is $F/2$. \textbf{Right:}
The graph displays the average fraction of time $\left\langle r\right\rangle $
[defined in Eq.~(\ref{eq:ratio})] spent by the particle in the second well, as a function of $\tau$.
When $\tau$ becomes large, the particle just jumps ballistically
over the second well and escape. Thus $\left\langle r\right\rangle $
goes to zero for large $\tau$. \label{fig:hopping}}

\end{figure}

We now illustrate the implications of Eq.~(\ref{eq:expansion large tau 2}) on escape processes. In particular, we show that an active particle can hop over a metastable state without ``feeling'' it. To see this, consider first the metastable potential $U(x)$ given in Fig.~\ref{fig:hopping}.
The potential is composed of two wells of equal depth, one for $x<x_{1}$,
and another one for $x_{1}<x<x_{2}$. The maximal force to overcome in order
to leave the first well is $U'(x_{\cri})=F$, whereas the maximal force
to overcome  the second well is $\frac{F}{2}$. We look at
the escape $x>x_{2}$ of an active AOUP particle starting in the first
well at $x=0$ with zero velocity. 

For small values of $\tau$, a velocity fluctuation $v>F$ brings
the particle out of the first well, and then, it spends some time
in the second well waiting for another velocity fluctuation of size
$v>\frac{F}{2}$ to escape. For large values of $\tau$ on the contrary,
the particle can use the velocity fluctuation $v>F$ created to leave
the first well in order to directly escape over the second well, without further
velocity fluctuations (see Supplementary movies). This behavior, as
we emphasize, is a consequence of the flatness of the
quasipotential for $x>x_{\cri}$ in the large $\tau$ limit, which means
that the mean escape time to reach $x>x_{2}$ is in fact the same
as to reach $x>x_{1}$. The fraction of time spent in the second well
$x_{1}<x<x_{2}$ vanishes in the limit $\tau\rightarrow+\infty$.
To show this latter point, we implemented a numerical simulation where we record the total times $T_{1}$ and
$T_{2}$ spent by each particle in the first and second well respectively,
and we numerically compute the ratio 
\begin{equation}
\left\langle r\right\rangle =\left\langle \frac{T_{2}}{T_{1}+T_{2}}\right\rangle, \label{eq:ratio}
\end{equation}
where the average is done over escape events, for each value of $\tau$.
Simple scaling arguments show that the large $\tau$ behavior of $\left\langle r\right\rangle $
is
\[
\left\langle r(\tau)\right\rangle \underset{\tau\rightarrow+\infty}{\propto}\frac{\log(\tau)}{\tau}\,.
\]
 We show in Fig.~\ref{fig:hopping} that the function $\left\langle r\right\rangle $
indeed decreases slowly with $\tau$. In addition, we illustrate the result with the two supplementary movies available at \cite{supplement}.

\section{AOUP in the small $\tau$ limit\label{sec:AOUP-in-the}}

We now detail the full calculation of the quasipotential in the small $\tau$ limit. Here we use a Lagrangian approach described in Sec.~\ref{subsec:Summary-of-the}, that allows for a generic organization of the perturbation expansion. The Lagrangian is given in Eq.~(\ref{eq:LagLag2}). Recall that the quasipotential at a point $x\in\mathcal{B}_0$ is obtained by evaluating the action on the `instanton trajectory' whose solution is given through the saddle-point equation
\begin{align}
\label{eq:eqvarsmalltau}
\ddot{x}-U'(x) U''(x)
-
\tau ^2 
  \Big(
\ddddot x +\dot{x}^3 U^{(4)}(x)-\ddot{x} \left(U''(x)^2-3 \dot{x} U'''(x)\right)-\dot{x}^2 U'''(x) U''(x)
  \Big)
=0
\,.
\end{align}
Assuming that $x_0$ is the location of the optimal minimum (the generalization being straightforward using standard methods, see for example~\cite{baek2015singularities}) the optimal path $x(t)$ starts from the local minimum $x_0$ of $U$ in $\mathcal{B}_0$  at time $t=-\infty$ and arrives in $x$ at time $t=0$. The Hamiltonian boundary condition~(\ref{eq:BC1}) on the momentum $P_v$ vanishing at $t=0$ implies that $\dot x(0)=0$. The boundary conditions on Eq.~(\ref{eq:eqvarsmalltau}) are given in Eq.~(\ref{eq:BC}).
Finally, note that conservation of `energy' [see Eqs.~(\ref{eq:hamiltonian}) and~(\ref{eq:Hconservation})] enforces the optimal path to satisfy
\begin{equation}
  \label{eq:csvE0}
H(x,\dot x,\ddot x, \dddot x)  \equiv -U'(x)^2+\tau ^2 \left(-2 \dot{x}^3 U'''(x)+\dot{x}^2 U''(x)^2+\ddot{x}^2-2 \dot{x} \dddot{x}\right)+\dot{x}^2  = 0
\,.
\end{equation}

\subsection{Expansion order by order}
\label{sec:expansionorderbyorder}

As shown in Sec.~\ref{sec:harmonic} (see Fig.~\ref{fig:harmonic fluctuations}) for the harmonic case, the instanton solution $x(t)$ presents two distinct regimes of time in the small $\tau$ limit: (i) a `bulk' regime $t\ll -\tau$ where the solution is uniformly close to the $\tau=0$ instanton, and where an expansion in powers of $\tau$ is well defined; and (ii) a `boundary-layer' regime $t\sim -\tau$ where the instanton cannot be obtained as a small perturbation around the $\tau=0$ instanton solution. One therefore has to treat the boundary layer using a singular perturbation theory. To this end, in what follows we determine order by order in powers of $\tau$ the form of the instanton in the bulk and in the boundary layer; these forms are determined up to constants that we obtain using standard asymptotic matching techniques. 


Namely, we write the instanton solution in the bulk regime as
\begin{equation}
  \label{eq:devbulk}
  x(t) = x_{\bu}(t) \equiv x^{(0)}(t) + \tau\, x^{(1)}(t) + \tau^2 x^{(2)}(t) + \tau^3 x^{(3)}(t) + \ldots \,,
\end{equation}
with the boundary condition $x_{\bu}(-\infty)=x_0$.  In the boundary layer, following standard approaches, we introduce $x(t)= x_{\rm{bl}}(t/\tau)$ and defines $t' = t/\tau$. The solution in this regime is expanded as
\begin{equation}
  \label{eq:devBLsmalltau}
  x_{\rm{bl}}(t')=  x^{(0)}_{\rm{bl}}(t') + \tau\, x^{(1)}_{\rm{bl}}(t') + \tau^2 x^{(2)}_{\rm{bl}}(t') + \tau^3 x^{(3)}_{\rm{bl}}(t') + \ldots \,,
\end{equation}
and Eq.~(\ref{eq:BC}) implies that $x_{\rm{bl}}(0)=0$ and $\dot{x}_{\rm{bl}}(0)=0$.
The bulk and boundary solution can be asymptotically matched by demanding that $x_{\bu}(t)\big|_{t\to 0}$ and $x_{\rm{bl}}(t/\tau)\big|_{t\to-\infty}$ have the same functional form.
The overall structure of the perturbation theory is implemented by enforcing the zero-energy condition~(\ref{eq:csvE0}) order by order in $\tau$ for both expansions for the bulk~(\ref{eq:devbulk}) and the boundary layer~(\ref{eq:devBLsmalltau}). 

In what follows we specify the solution order by order.

\subsubsection{Zeroth order}

\hspace*{\parindent}%
In the bulk regime, using~(\ref{eq:devbulk}) with~(\ref{eq:csvE0}) one finds at zeroth order in $\tau$ that
\begin{equation*}
  \big(U'(x^{(0)})\big)^2-\big(\dot{x}^{(0)}\big)^2=0
\,,
\end{equation*}
One thus sees that $x^{(0)}(t)$ is equal to the equilibrium instanton $\tilde{x}(t)$, which is given by the solution of
\begin{equation}
\label{eq:xtilde0t}
\dot{\tilde{x}}(t)  =U'(\tilde{x}(t))
\qquad
\text{with}
\quad
\tilde x(-\infty) = x_0 
\ \text{ and } \
\tilde x(0) = x 
\,.
\end{equation}
Next, we express higher orders of the bulk expansion~(\ref{eq:devbulk}) as a function of $\tilde x(t)$.

In the boundary-layer regime, using Eqs.~(\ref{eq:csvE0}) and~(\ref{eq:devBLsmalltau}) one finds to zeroth order in $\tau$ that
\begin{equation*}
2 \dot{x}^{(0)}_{\rm{bl}} \, \dddot{x}^{(0)}_{\rm{bl}}
  -\big(\ddot{x}^{(0)}_{\rm{bl}}\big)^2
  -\big(\dot{x}^{(0)}_{\rm{bl}}\big)^2
=0
\,,
\end{equation*}
whose solution is
\begin{equation*}
 x^{(0)}_{\rm{bl}}( t' ) =   \Big(2 C_1 t'-C_1^2 e^{-t'}+e^{t'}\Big)C_2+C_3 
\,,
\end{equation*}
with $C_k$'s constants. The asymptotic matching condition implies that $x^{(0)}_{\rm{bl}}(t')$ has to remain constant as $ t'\to -\infty$. This imposes $C_2=0$, which with the boundary condition at $t'=0$ implies $C_3=x$. Therefore, $ x^{(0)}_{\rm{bl}}( t' ) =  x$.

\medskip
The quasipotential is obtained by decomposing the time interval $]-\infty,0]$ into $]-\infty,- A\tau]\: \cup \: ]-A\tau,0] $ and using the bulk and boundary-layer expressions of the instanton trajectory:
\begin{equation}
\label{eq:decompquasipotintegral}
\Phi(x)
=
  \int_{-\infty}^{-A \tau} \mathcal L (x_{\bu},\dot x_{\bu}) \, \dd t
+
  \int_{-A}^0 \tau \, \mathcal L (x_{\rm{bl}}, \tau  \dot{x}_{\rm{bl}}) \, \dd t'
\,,
\end{equation}
Clearly, for this to be consistent the result should be independent of $A$ in the large $A$ limit. Using the solution we have obtained one finds, as expected, that the boundary layer does not contribute at order $\tau^0$ and one recovers the equilibrium result
\begin{equation*}
 \Phi(x) = U(x)-U(x_0) + O(\tau) 
 \,.
\end{equation*}

\subsubsection{First order}

\hspace*{\the\parindent}%
In the bulk regime, using~(\ref{eq:csvE0}) and~(\ref{eq:devbulk}) and taking into account zeroth order results, one finds that
\begin{equation*}
 \dot{\tilde{x}} \dot{x}^{(1)}  - \ddot{\tilde{x}} x^{(1)}
 = 0
\,,
\end{equation*}
where we also used Eq.~(\ref{eq:xtilde0t}) for the equilibrium instanton $\tilde x(t)$, to eliminate derivatives of $U$. The solution is then
\begin{equation}
  \label{eq:solx1}
 x^{(1)} (t) = c_1 \dot{\tilde{x}}(t)\,.
\end{equation}

In the boundary-layer regime, Eqs.~(\ref{eq:eqvarsmalltau}) and~(\ref{eq:devBLsmalltau}) give
\begin{equation*}
\ddddot{x}^{(1)} _{\rm{bl}}
-
  \ddot{x}^{(1)} _{\rm{bl}}
=0 \,,
\end{equation*}
whose solution, satisfying the boundary conditions\footnote{%
Note that since $x_{\rm{bl}}(0)=x$ the boundary condition on the higher order terms in $t'=0$ are $ x^{(k)} _{\rm{bl}}(0)=0$ and $\dot{x}^{(k)} _{\rm{bl}}(0)=0$ ($k\geq 1$).
} at $t'=0$ is
\begin{equation*}
 x^{(1)} _{\rm{bl}}(t')
=
D_1 \big(e^{t'}-1-t'\big)+D_2 \big(e^{-t'}-1+t'\big)
\end{equation*}
and the zero-energy condition, together with the solution at zeroth order, imposes $(D_1+D_2){}^2-U'(x)^2=0$.

The asymptotic matching then imposes $D_2=0$ so that $D_1=\pm U'(x)$.
To determine the sign of $D_1$ and the value of $c_1$ in~(\ref{eq:solx1}) one compares the asymptotic behaviors of the bulk and the boundary-layer:
\begin{align*}
   x_{\bu}( t)\big|_{t=\tau t'} 
& 
\
\overset{\phantom{t'\to-\infty}}{
 \underset{t\to 0}{\approx}
 }
\
 x+   \big[ c_1 U'(x)+t'\, U'(x)\big] \tau 
+ \ldots
\\[1mm]
  x_{\rm{bl}}(t') 
&
\
 \underset{ t'\to-\infty}{\approx}
\
  x - \big[D_1\,(1+t')\big]\tau
+ \ldots
\end{align*}
(Note that for $x_{\bu}(t)$ both $x^{(0)}(t)$ and $x^{(1)}(t)$ contribute, and we used the equation of motion $\dot{\tilde{x}}(t)  =U'(\tilde{x}(t))$ to evaluate $\dot{\tilde{x}}(0)  =U'(x)$.) The asymptotic matching therefore imposes $D_1=-U'(x)$ and $c_1=1$.
To summarize, the first-order perturbations in the bulk and boundary layer are:
\begin{equation}
\label{eq:resultperturborder1}
   x^{(1)}(t) =  \dot{\tilde{x}}(t) \ ; 
   \qquad
  x^{(1)} _{\rm{bl}}(t') = \big(e^{t'}-1-t'\big) \, U'(x) \,.
\end{equation}
We remark that at this first order, the boundary layer $x_{\bl}(t/\tau)$ contains terms proportional to $ \tau e^{t/\tau}$ which are non-perturbative in~$\tau$ as $\tau\to 0$ at fixed~$t$. This fact explains how the boundary layer contributes to the expression of the quasipotential in $\Phi(x)=\int_{-\infty}^{0}  \mathcal L(x(t),\dot x(t))\, \dd t$ even though it only affects this time integral over a duration~$\tau$.

\medskip

As we discussed before Eq.~(\ref{eq:decompquasipotintegral}) the quasipotential is given by
\begin{align}
\label{eq:decompquasipotintegral2}
\Phi(x)
&=
\lim_{A\to\infty}
\Big[
\Phi^A_{\bu}(x)
+
\Phi^A_{\bl}(x)
\Big]
\end{align}
with%
\begin{align}          
\label{eq:defPhiAbubl}
\Phi^A_{\bu}(x)
=
  \int_{-\infty}^{-A \tau} \mathcal L (x_{\bu},\dot x_{\bu}) \, \dd t 
\qquad\text{and}\qquad
\Phi^A_{\bl}(x)
=
  \int_{-A}^0\tau \, \mathcal L (x_{\rm{bl}}, \tau  \dot{x}_{\rm{bl}}) \, \dd t' \,.
\end{align}
Using the solution for $x_{\bu}(t)$ one finds to order $\tau$
\begin{equation*}
  \mathcal L (x_{\bu},\dot x_{\bu})
  =
  \dot{\tilde{x}}\,U'(\tilde{x})+4 \tau  \dot{\tilde{x}} \ddot{\tilde{x}}
  =
  \partial_t \Big[U(\tilde{x})+2 \tau  \dot{\tilde{x}}^2 \Big]
,
\end{equation*}
so that
\begin{equation*}
  \Phi^A_{\bu}(x)
  = 
  \Big[U(\tilde{x})+2 \tau  \dot{\tilde{x}}^2 \Big]_{-\infty}^{-A\tau}
  = 
  \Big[U(\tilde{x})+2 \tau  U'({\tilde{x}})^2 \Big]_{-\infty}^{-A\tau}
\,.
\end{equation*}
Using then $U'(\tilde x(-\infty))=U'(x_0)=0$ and $\tilde x(-A\tau)=\tilde x(0)-A\tau \dot{\tilde x}(0)=\tilde x(0)-A\tau U'(x)$ one finally has
\begin{equation}
\label{eq:resPhiAbuo1}
  \Phi^A_{\bu}(x)
  = 
  U(x)-U(x_0)
  +
  2 \tau (-A+1)
\,   U'(x)^2
  + O(\tau^2)
\,.
\end{equation}
In the boundary layer one finds, up to order $\tau$
\begin{equation*}
  \tau \,  \mathcal L (x_{\rm{bl}}, \tau  \dot{x}_{\rm{bl}})
  =
  \tau\, \big(e^{t'}-1\big)^2   U'(x)^2
  +O(\tau^2)
\end{equation*}
so that Eq.~(\ref{eq:defPhiAbubl}) gives
\begin{equation}
\label{eq:resPhiAblo1}
  \Phi^A_{\bl}(x) = \frac 12 \tau\, \Big[ 2 A-3-e^{-2 A}+4 e^{-A}\Big] U'(x)^2
  +O(\tau^2)
\,.
\end{equation}
Adding~(\ref{eq:resPhiAbuo1}) and~(\ref{eq:resPhiAblo1}) one sees that the terms linear in $A$ compensate as we take $A \to \infty$. 
We finally obtain from~(\ref{eq:decompquasipotintegral2}) the quasipotential to order $\tau$
\begin{equation}
  \label{eq:devqpo1X}
  \Phi(x) = U(x)-U(x_0) + \frac 12 \tau\,U'(x)^2  + O(\tau^2) 
\,.
\end{equation}

\medskip
\noindent\underline{Remark}:
If we had ignored the boundary layer, we would obtain that $c_1=0$ in~(\ref{eq:solx1}) so as to satisfy the boundary condition~(\ref{eq:BC}) in $t=0$ [because $\dot {\tilde x}(0)=U'(x)\neq 0$ in general], meaning that the instanton $x(t)$ would present no correction of order $\tau$.
This would imply that the only contribution of order $\tau$ to the quasipotential is given by the double-product in the Lagrangian~(\ref{eq:LagLag2}), that is:
\begin{equation*}
   \frac 14 \tau \int \partial_t\Big[ \big(\dot{\tilde x}+U'(\tilde x)\big)^2\Big] \dd t = 
   \frac 14 \tau \Big[ \big(\dot{\tilde x}+U'(\tilde x)\big)^2\Big]_{-\infty}^0
   = \tau \, U'(x)^2
\end{equation*}
which is wrong by a factor $\frac 12$ compared to the actual result~(\ref{eq:devqpo1X}).
This illustrates how the presence of the boundary layer affects the determination of the quasipotential: it modifies the boundary condition in $t\to0$ seen by the instanton, and  it contributes at order $\tau$ (and higher orders) to  the integral of the Lagrangian in the action.

\subsubsection{Second order}

\hspace*{\the\parindent}%
In the bulk regime, inserting Eq.~(\ref{eq:devbulk}) in the zero-energy condition~(\ref{eq:csvE0}) and taking into account the result obtained at zeroth and first order, one finds that the second-order correction to the bulk instanton is given by the solution of
\begin{equation*}
 \dot{\tilde{x}} 
\dot{x}^{(2)}
-
 \ddot{\tilde{x}}
x^{(2)}
 +
2 (\ddot{\tilde{x}}^2-\dot{\tilde{x}} \dddot{\tilde{x}})
=
0 \,.
\end{equation*}
Here we used Eq.~(\ref{eq:xtilde0t}) to eliminate occurrences of $U(\tilde x)$ and its derivatives.
The solution to this equation is given by
\begin{equation*}
 x^{(2)}(t) = 
d_1\, \dot{\tilde{x}}(t)+\frac{5 }{2}\, \ddot{\tilde{x}}(t)
\,,
\end{equation*}
where $d_1$ is a constant. In the boundary-layer regime, inserting Eq.~(\ref{eq:devBLsmalltau}) in~(\ref{eq:eqvarsmalltau}) and using the results obtained at zeroth and first order one finds in a similar manner 
\begin{equation*}
\ddddot{x}\,^{(2)}_{\rm{bl}}-\ddot{x}^{(2)}_{\rm{bl}}+U'(x) U''(x) 
=0
\,.
\end{equation*}
Using the boundary conditions $x^{(2)}_{\rm{bl}}(0)=0$ and $\dot{x}^{(2)}_{\rm{bl}}(0)=0$ with the zero-energy constraint and demanding that the solution does not diverge exponentially as $t\to-\infty$, so that the bulk solution can be matched, gives
\begin{equation}
  \label{eq:xh2sol}
x^{(2)}_{\rm{bl}}(t')
=
\Big(
 1 +
t'
+
\frac{t'^2}{2}
-e^{t'}
\Big)
U'(x) U''(x)
\,.
\end{equation}
Then using asymptotic matching one finds $d_1=-\frac{3}{2} U''(x)$ so that finally
\begin{equation}
  \label{eq:solx2}
x^{(2)}_{\bu}(t) = 
-\frac{3}{2} U''(x)\,
 \dot{\tilde{x}}(t)+\frac{5 }{2}\, \ddot{\tilde{x}}(t)
\,.
\end{equation}

We now proceed to evaluate the quasipotential. The order $\tau$ calculation of the quasipotential gave a local function of the potential, see Eq.~(\ref{eq:devqpo1X}). We now show that at order $\tau^2$ the quasipotential, as stressed in the introduction, becomes non-local.

We saw that the solution to order $\tau$ for the functions $\Phi^A_{\bu}(x)$ and $\Phi^A_{\bl}(x)$ had divergent contributions at order $\tau$ when taking the $A\to\infty$ limit which compensated each other when $\Phi^A_{\bu}(x)$ and $\Phi^A_{\bl}(x)$ were summed.
Performing a similar analysis at order $\tau^2$ (and higher) is more complex. Here we avoid this procedure. To do so we replace $\mathcal L(x,\dot x)$ in the integrals of Eq.~(\ref{eq:defPhiAbubl}) defining  $\Phi^A_{\bu}(x)$ and $\Phi^A_{\bl}(x)$ by
\begin{equation}
  \label{eq:newL}
  \mathcal L_F(x,\dot x)
=
  \mathcal L(x,\dot x) - \,\partial_t\big[\underbrace{F_0(x) + \tau \,F_1(x)+\ldots}_{\equiv F(x)}\big] \,.
\end{equation}
Namely, we remove a counter-term that is a total derivative. In the action, the counter-term becomes a boundary term that can be evaluated separately, before any expansion (and before splitting $]-\infty,0]$ into a bulk and boundary-layer regime).
In sum, we now write
\begin{align}
\Phi(x)
&=
 \big[F(x(t))\big]_{x_0}^x
\
+
\lim_{A\to\infty}
\Big[
\Phi^A_{\bu}(x)
+
\Phi^A_{\bl}(x)
\Big]
,
\label{eq:decompAF}
\end{align}
with%
\begin{align}          
\label{eq:defPhiAbublF}
\Phi^A_{\bu}(x)
=
  \int_{-\infty}^{-A \tau} \mathcal L_F (x_{\bu},\dot x_{\bu}) \, \dd t
\qquad\text{and}\qquad
\Phi^A_{\bl}(x)
=
  \int_{-A}^0\tau \, \mathcal L_F (x_{\rm{bl}}, \tau  \dot{x}_{\rm{bl}}) \,\dd t'
\,.
\end{align}
The functions $F_0(x)$ and $F_1(x)$ in Eq.~(\ref{eq:newL}) are taken so that asymptotic matching does not give any divergence as $A\to\infty$.
Taking
\begin{equation}
  F_0(x) = U(x) 
  \qquad\text{and}\qquad
  F_1(x) = U'(x)^2\,,
\label{eq:resF0F1}
\end{equation}
one finds after a straightforward expansion of the integrals in~(\ref{eq:defPhiAbublF}), separately, and using~(\ref{eq:decompAF}) that, as announced previously
\begin{equation}
  \label{eq:devqpo2X}
  \Phi(x) = U(x)-U(x_0) 
  + \frac 12 \tau\,U'(x)^2  
-\frac{1}{2}\tau^2
\int_{x_{0}}^{x}\big[U'(y)\big]^2U'''(y)  \:\dd y
  + O(\tau^3)\,.
\end{equation}

For instance, the choice of function for $F_0(x)$ avoids the appearance of terms linear in $A$ in the bulk and boundary-layer expressions~(\ref{eq:resPhiAbuo1}) and~(\ref{eq:resPhiAblo1}) to order $\tau$.
At order $\tau^2$ one finds by inserting the series~(\ref{eq:devBLsmalltau}) into~(\ref{eq:defPhiAbublF}) that the contribution of order $\tau^2$ to the integrand of~(\ref{eq:defPhiAbublF}) is
\begin{align}
\tau \, \mathcal L_F (x_{\rm{bl}}, \tau  \dot{x}_{\rm{bl}})  \: \big|_{\tau^2}
=& \:
\Big\{ 
 3 U'(x)^2 U''(x) \: e^{2 t'} 
 + U'(x) \Big[F_1'(x)-\left(t'+5\right) U'(x) U''(x)\Big] e^{t'}
\nonumber
\\
& \quad +U'(x) \Big[2 U'(x) U''(x)-F_1'(x)\Big]
\Big\} \tau^2\,.
  \label{eq:devPhiAFhat}
\end{align}
When performing the integral over $t'$ in~(\ref{eq:defPhiAbublF}), the second line in~(\ref{eq:devPhiAFhat}) yields terms which diverge with $A\to\infty$. 
The choice of function $F_1(x)$ in~Eq.~(\ref{eq:resF0F1}) then avoids this divergence.

\smallskip
We note that the result~(\ref{eq:devqpo2X}), derived here through a singular perturbation theory for the instanton, can also be obtained through a different approach. In an early work, K{\l}osek-Dygas, Matkowsky and Schuss~\cite{klosek-dygas_colored_1988} studied the escape problem using a Fokker--Planck approach (and in the small $\tau$ expansion). The approach yields the steady-state solution $P_{\text{st}}(x,\dot x)$ of the Fokker--Planck equation on the joint probability density $P(x,\dot x,t)$, within a large-deviation approach at small $D$. The authors obtain an expansion of $ \lim_{D\to 0} D \log P_{\text{st}}(x,\dot x)$ in powers of $\tau$ (their Eq.~(21a)). If one takes the marginal over the variable $\dot x$ in their results it is easy to check that the result is again~(\ref{eq:devqpo2X}). This alternative derivation however does not unveil the singular nature of the optimal trajectory.

\section{AOUP in the large $\tau$ limit\label{sec:AOUP-in-the-large}}
In this part it is more convenient to use the Hamiltonian formalism. The large $\tau$ limit is best understood by rescaling time according to $t\leftarrow t/\tau$.  After the time rescaling, the Hamiltonian
(\ref{eq:hamiltonian}) takes the form

\begin{equation}
H(x,v,P_{x},P_{v})=\tau\left(v-U'(x)\right)P_{x}-vP_{v}+P_{v}^{2}\,,
\label{eq:hamiltonian2}
\end{equation}
so that the instanton equations are
\begin{equation}
\begin{casesb}
\dot{x} & =\tau\left(v-U'(x)\right),\\
\dot{v} & =-v+2P_{v}\,,
\end{casesb}\label{eq:hamilton space2}
\end{equation}
and 
\begin{equation}
\begin{casesb}
\dot{P_{x}} & =\tau U''(x)P_{x}\,,\\
\dot{P_{v}} & =P_{v}-\tau P_{x}\,.
\end{casesb}\label{eq:hamilton momentum}
\end{equation}
For ease of presentation, we now split the discussion in the escape problem and the calculation of the general quasipotential. In both cases it will be useful to keep in mind a general potential of the form presented in Fig.~\ref{fig:large tau barrier}.

\subsection{The quasipotential for {\textnormal{$x > x_{\cri}$}}}
 As before we denote the position of the highest point of the potential by $x_{\sa}$ and the position of the (unique) inflection point where  $U''(x_{\cri})=0$ by $x_{\cri}$. To evaluate the quasipotential we look for an \emph{instanton path} by solving Eqs.~(\ref{eq:hamilton space2})
and~(\ref{eq:hamilton momentum}) with the boundary conditions
\begin{equation}
x(t),v(t)\underset{t\rightarrow-\infty}{\longrightarrow}0\;\; ; \;P_{x}(t),P_{v}(t)\underset{t\rightarrow -\infty}{\longrightarrow}0\label{eq:BC1b}
\,.
\end{equation}
The boundary conditions at later times are more subtle. Initially using the time translation invariance of the instanton path we choose
\begin{equation}
x(0)=x_{\cri} \;,\label{eq:inflexion crossing}
\end{equation}
but, as will become clear, we also have to specify boundary conditions at positive times. This will be carried out below.

\subsubsection{Bulk solution\label{subsec:Bulk-solution}}

We first show that $P_{x}=0$ to any order for $t$ not close to $t=0$, where as we will show, a boundary layer appears. Consider $\delta>0$, and the instanton path $x(t)$ on the interval
$]-\infty,-\delta]$: as $x(t)$ is a monotonic function of $t$,
it is clear that $x(t)<x(-\delta)<0$ on this part of the path. Moreover, since we assume the existence of a single inflection point, $U''$ is strictly positive in $[x_{0},x(-\delta)]$. Therefore, there
exists some constant $c_\delta>0$ such that 
\begin{equation}
\forall t<-\delta,\;U''(x(t))>c_\delta\, .\label{eq:bound second derivative}
\end{equation}
Then the first equation of~(\ref{eq:hamilton momentum}) shows that
$\dot{P_{x}}>c_\delta \tau P_{x}$ over the time interval $]-\infty,-\delta]$.
Thus, we deduce the bound
\[
\forall t<-\delta,\;\left|P_{x}(t)\right|<\left|P_{x}(-\delta)\right|e^{c_\delta \tau t} \,.
\]
The momentum $P_{x}$ relaxes faster than $e^{c_\delta\tau t}$ when $t\rightarrow-\infty$.
A similar argument shows that $P_{x}$ relaxes exponentially faster
that $e^{c_\delta'\tau t}$, with $c'_\delta<0$ when $t\rightarrow+\infty$. Therefore, since we are interested in the limit $\tau \to \infty$, these bounds prove
that an expansion of $P_{x}$ in powers of $\tau$ vanishes to any
order in $\tau$, except in the immediate vicinity of $x_{\cri}=0$.
This simple observation indicates the existence of a boundary layer
in the vicinity of $t=0$ that connects the part $]-\infty,0[$ to
the part $]0,+\infty[$ of the instanton path. The scaling with $\tau$
of the boundary layer is nontrivial, and will be derived in the following.\\

In this section, we solve Eqs.~(\ref{eq:hamilton space2}-\ref{eq:hamilton momentum})
in the bulk regimes $t<0$ and $t>0$. The boundary layer ($t$ close to zero) is solved in Sec.~\ref{sec:BLsolution}. With $P_{x}=0$, the instanton equations in
the bulk are
\begin{equation}
\begin{casesb}
\dot{x} & =\tau\left(v-U'(x)\right),\\
\dot{v} & =-v+2P_{v}\,,\\
\dot{P_{v}} & =P_{v}\,.
\end{casesb}\label{eq:bulk}
\end{equation}
To solve the system~(\ref{eq:bulk}), we expand the solution as
\begin{equation}
P_{v}(t) =P_{v}^{(0)}(t)+\frac{1}{\tau^{n}}P_{v}^{(1)}(t)+\frac{1}{\tau^{2n}}P_{v}^{(2)}(t)+...\;,
\label{eq:expansion bulk}
\end{equation}
and similarly for $x(t)$ and $v(t)$. The exponent $n$
is related to the size of the boundary layer close to $t=0$, and
has to be determined self-consistently. However, before considering the corrections we first solve the zeroth order of the problem.

\medskip
\noindent{\bf Zeroth order}: To this order we have, using Eq.~(\ref{eq:bulk})
\begin{equation}
\begin{casesb}
U'(x^{(0)}) & =v^{(0)},\\
\dot{v}^{(0)} & =-v^{(0)}+2P_{v}^{(0)},\\
\dot{P}_{v}^{(0)} & =P_{v}^{(0)}.
\end{casesb}\label{eq:bulk-0}
\end{equation}
The solution for $P_{v}^{(0)}$ and $v^{(0)}$ is straightforward.
Taking into account the boundary conditions~(\ref{eq:BC1b}) in $t =- \infty$,
we get
\begin{equation}
t<0:\begin{casesb}
v^{(0)}(t) & =A_{0}e^{t}\,,\\
P_{v}^{(0)}(t) & =A_{0}e^{t}
\end{casesb}
\qquad \text{ and }\qquad 
t>0:\begin{casesb}
v^{(0)}(t) & =B_{0}e^{-t},\\
P_{v}^{(0)}(t) & =0.
\end{casesb}\label{eq:solution0th}
\end{equation}
where $A_{0}>0$. Since the third equation in~(\ref{eq:bulk-0}) implies that $P_v$ is either zero identically or diverges to infinity for $t>0$ one has to take $P_v=0$ as expected. Beyond $x_{\cri}$ the motion of the instanton follows a noiseless relaxation path. Therefore its contribution to the action is zero. Then the solution for $v^{(0)}$ follows with $B_0>0$.

The solution for $x^{(0)}(t)$ has to be computed
from the first equation of~(\ref{eq:bulk-0}), inverting the function
$U'$. In the vicinity of $x_{\cri}=0$, we can expand $U'$ as
\[
U'(x)=\beta-\gamma x^{2}+O(x^{3}) \,.
\]
Then using $U'(x^{(0)})=v^{(0)},$ and the solution~(\ref{eq:solution0th})
we get
\[
x^{(0)}(t)\underset{t\rightarrow0^{-}}{\sim}\sqrt{\frac{1}{\gamma}\left(\beta-A_{0}e^{t}\right)} \,.
\]
Note that since the instanton path for $t>0$ does not contribute to the action we do not have to consider the solution in that regime. The constraint $x(0)=0$ imposes $A_{0}=\beta$ and thus
\begin{equation}
x^{(0)}(t)\underset{t\rightarrow0^{-}}{\sim}-\sqrt{\frac{\beta}{\gamma}\left|t\right|}\label{eq:eqt0}
\end{equation}
Eq.~(\ref{eq:eqt0}) gives the asymptotic behavior
that the boundary layer solution has to satisfy. As we show just below,
it also constrains the size of the boundary layer.

\medskip
\noindent {\bf Boundary layer scaling}: In the vicinity of $x_{\cri}=0$, the first equation of~(\ref{eq:hamilton momentum})
gives
\begin{equation}
\dot{P_{x}}=-2\tau\gamma xP_{x}\,.\label{eq:momentumPx} 
\end{equation}
Plugging the solution~(\ref{eq:eqt0}) into~(\ref{eq:momentumPx})
gives
\[
\dot{P_{x}}=-2\tau\sqrt{\gamma\beta\left|t\right|}P_{x}\,,
\]
which can be integrated to give
\[
P_{x}(t)\propto e^{-2\sqrt{\gamma\beta}\frac{2\tau}{3}\left|t\right|^{3/2}}.
\]
This proves that $P_{x}$ relaxes over a typical time scaling $\propto\frac{1}{\tau^{2/3}}$.  
The boundary layer scaling variable is also $\tau^{2/3}t$, as we show below. In what follows it will become clear that this implies that for the expansion
(\ref{eq:expansion bulk}) to be consistent one has to take $n=\frac{1}{3}$.

\medskip
\noindent {\bf Higher orders}: Using the linearity of the bulk equations for $v$ and $P_{v}$, we
can express the solution to any order $k$ as
\begin{equation}
t<0:\begin{casesb}
v^{(k)}(t) & =A_{k}e^{t}\,,\\
P_{v}^{(k)}(t) & =A_{k}e^{t}
\end{casesb}
\qquad \text{ and }\qquad
t>0:\begin{casesb}
v^{(k)}(t) & =B_{k}e^{-t}\,,\\
P_{v}^{(k)}(t) & =0\,.
\end{casesb}\label{eq:solutionNth}
\end{equation}
One can then find the solution $x^{(k)}(t)$ using the expansion of $\dot{x}=\tau\left(v-U'(x)\right)$
to $k^{th}$ order in $\frac{1}{\tau^{1/3}}$. The constants $A_{k}$
and $B_{k}$ have to be determined by matching the bulk solution to
the boundary layer as we do below to up to order two. We now turn to consider the solution in the boundary layer in the vicinity of $x_{\cri}$.

\subsubsection{Boundary layer solution}
\label{sec:BLsolution}

To find the boundary layer equations, we take Eqs.~(\ref{eq:hamilton space2}-\ref{eq:hamilton momentum})
and again use the expansion of $U'(x)$ to second order near $x_{\cri}=0$
\[
U'(x)=\beta-\gamma x^{2}+O(x^{3})\,.
\]
This gives
\begin{equation}
\begin{casesb}
\dot{x} & =\tau\left(v-\beta+\gamma x^{2}\right),\\
\dot{v} & =-v+2P_{v}\,,
\end{casesb}\label{eq:hamilton space-1}
\end{equation}
and 
\begin{equation}
\begin{casesb}
\dot{P_{x}} & =-2\tau\gamma xP_{x}\,,\\
\dot{P_{v}} & =P_{v}-\tau P_{x}\,.
\end{casesb}\label{eq:hamilton momentum-1}
\end{equation}
Using the two parameters $\beta$ and $\gamma$, we can rescale all
fields in order to arrive at non-dimensional equations. We set $\tau'=\sqrt{\gamma\beta}\tau$,
$v'=v/\beta$, $P_{v}'=P_{v}/\beta$ , $x'=x\sqrt{\frac{\gamma}{\beta}}$
and $P_{x}'=\tau P_{x}/\beta$. We get
\begin{equation}
\begin{casesb}
\dot{x}' & =\tau'\left(v'-1+x'^{2}\right),\\
\dot{v}' & =-v'+2P_{v}'\,,
\end{casesb}\label{eq:hamilton space-1-1}
\end{equation}
and 
\begin{equation}
\begin{casesb}
\dot{P_{x}}' & =-2\tau'x'P_{x}'\,,\\
\dot{P_{v}}' & =P_{v}'-P_{x}'\,.
\end{casesb}\label{eq:hamilton momentum-1-1}
\end{equation}
Finally, we use the scaling of the boundary layer and we rescale time
according to $t'=\tau'^{\,2/3}t$ and for self-consistency define the variable $x''=x'\tau'^{\,1/3}$
and $P_{x}''=P_{x}'/\tau'^{\,2/3}$. Omitting the primes, the boundary
layer equations are
\begin{equation}
\begin{casesb}
\dot{x} & =\tau^{2/3}\left(v-1\right)+x^{2}\,,\\
\dot{v} & =-\frac{1}{\tau^{2/3}}v+\frac{2}{\tau^{2/3}}P_{v}\,,
\end{casesb}\label{eq:BL1}
\end{equation}
and 
\begin{equation}
\begin{casesb}
\dot{P_{x}} & =-2xP_{x}\,,\\
\dot{P_{v}} & =\frac{1}{\tau^{2/3}}P_{v}-P_{x}\,.
\end{casesb}\label{eq:BL2}
\end{equation}

We then expand the boundary layer solution as
\[
x_{\bl}(t)=x_{\bl}^{(0)}(t)+\frac{1}{\tau^{1/3}}x_{\bl}^{(1)}(t)+\frac{1}{\tau^{2/3}}x_{\bl}^{(2)}(t)+...
\]
and similar expansions for $P_{x,{\bl}}$, $P_{v,{\bl}}$, $ v_{\bl}(t)$. We now turn to solve these equation order by order.

\medskip
\noindent {\bf Leading order}: In this case the equations are dominated by terms of order $\tau^{2/3}$. This gives 
\[
v_{\bl}^{(0)}=1\,,
\]
 which is consistent with the bulk zeroth order solution found in
(\ref{eq:solution0th}) with $A_{0}=\beta$ (see the discussion before Eq.~(\ref{eq:eqt0})). This also implies $B_{0}=\beta$. It is then easy to see that the next term (of order $\tau^{1/3}$) vanishes:
\[
v_{\bl}^{(1)}=0\,,
\]
which implies $A_{1}=0$ and $B_{1}=0$ in the bulk, and that all
terms of order $\frac{1}{\tau^{1/3}}$ vanish in the bulk solution
(\ref{eq:solutionNth}).

\medskip

\begin{figure}
\begin{centering}
\includegraphics[width=7cm]{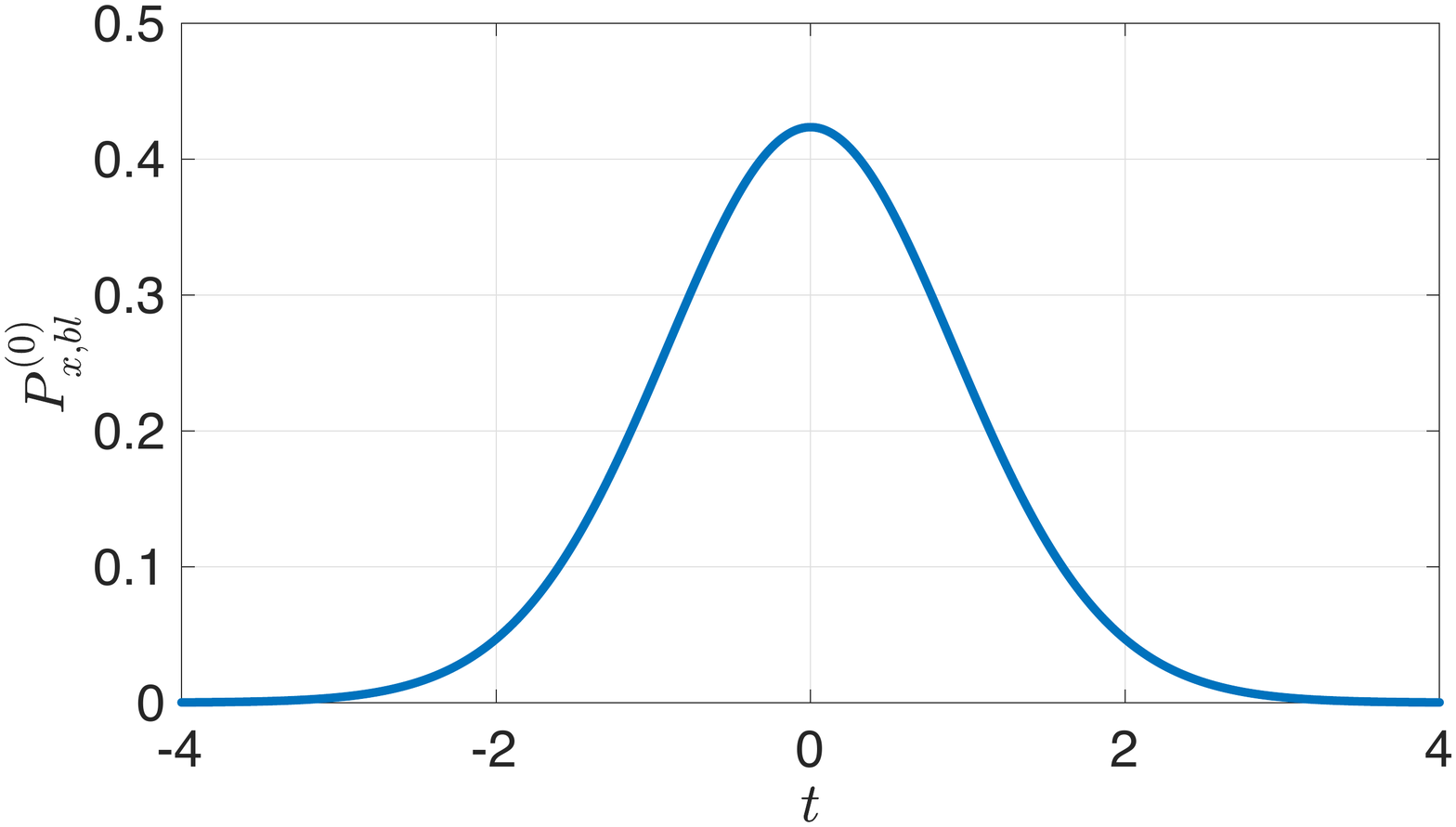}\includegraphics[width=7cm]{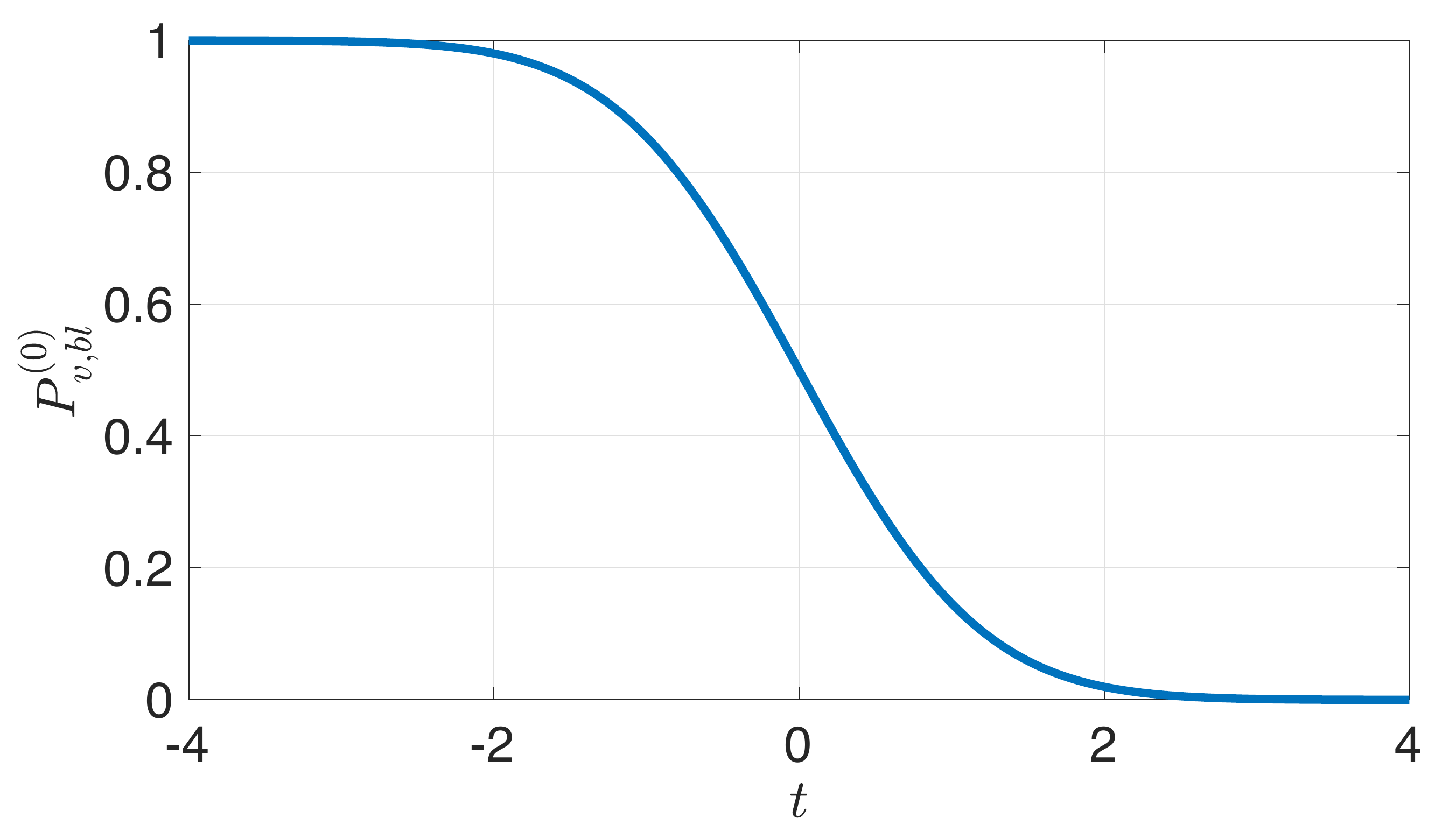}
\par\end{centering}
\begin{centering}
\includegraphics[width=7cm]{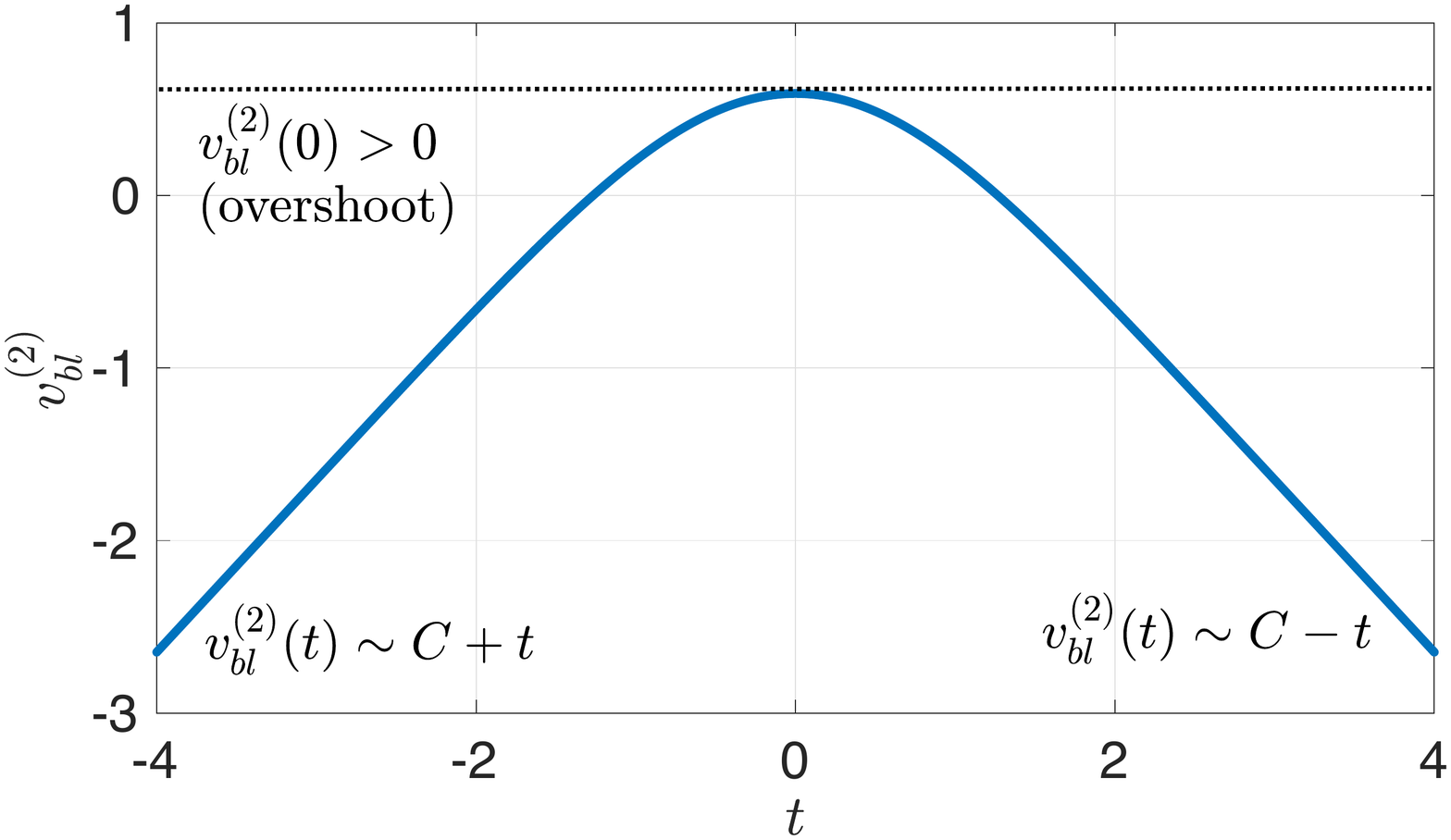}\includegraphics[width=7cm]{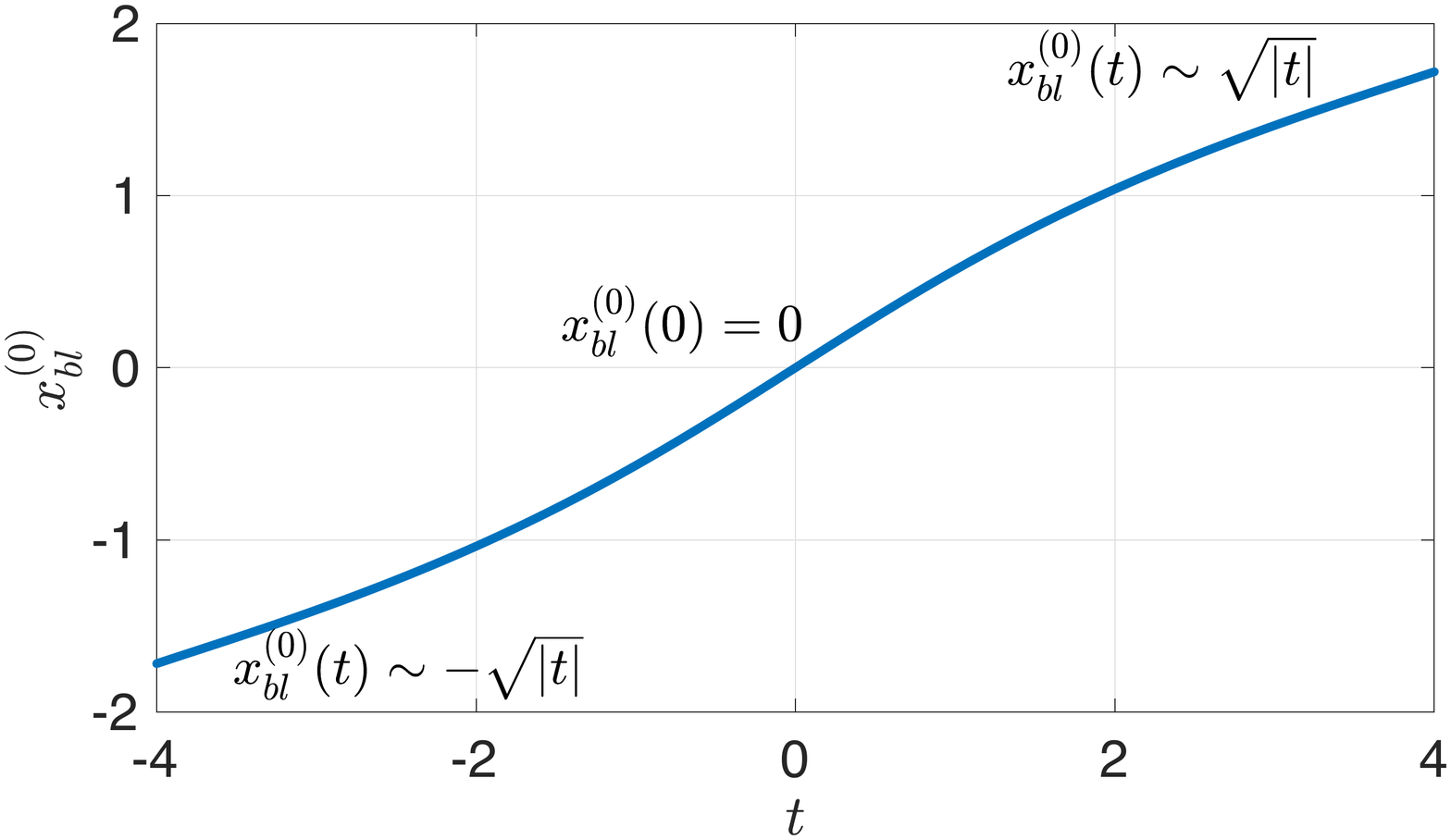}
\par\end{centering}
\caption{Numerical resolution of the undimensioned boundary layer equations Eqs.~(\ref{eq:BL1}-\ref{eq:BL2}). 
\label{fig:solution}}
\end{figure}

\noindent {\bf Zeroth order}: To next order the equation are
\begin{equation}
\begin{casesb}
\dot{x}_{\bl}^{(0)} & =v_{\bl}^{(2)}+\left(x_{\bl}^{(0)}\right)^{2},\\
\dot{v}_{\bl}^{(2)} & =-1+2P_{v,\bl}^{(0)}\,,
\end{casesb}\label{eq:BL1-1}
\end{equation}
and 
\begin{equation}
\begin{casesb}
\dot{P}_{x,\bl}^{(0)} & =-2x_{\bl}^{(0)}P_{x,\bl}^{(0)}\,,\\
\dot{P}_{v,\bl}^{(0)} & =-P_{x,\bl}^{(0)}\,.
\end{casesb}\label{eq:BL2-1}
\end{equation}
The boundary conditions for the set of Eqs.~(\ref{eq:BL1-1}-\ref{eq:BL2-1})
can be found by matching them with the bulk solution. The fact that $P_{x}=0$
to any order in the bulk gives
\begin{equation}
P_{x,\bl}^{(0)}\underset{t\rightarrow\pm\infty}{\longrightarrow}0\,.\label{eq:match1}
\end{equation}
The matching with the bulk solution~(\ref{eq:solution0th}) with $A_{0}=B_{0}=\beta$
gives 
\begin{equation}
\begin{casesb}
P_{v,\bl}^{(0)} & \underset{t\rightarrow-\infty}{\longrightarrow}1\,,\\
P_{v,\bl}^{(0)} & \underset{t\rightarrow+\infty}{\longrightarrow}0\,.
\end{casesb}\label{eq:match2}
\end{equation}
The bulk estimate~(\ref{eq:eqt0}) close to $t=0$ gives
\begin{equation}
x_{\bl}^{(0)}(t)\underset{t\rightarrow-\infty}{\sim}-\sqrt{\left|t\right|}\,,\label{eq:match3}
\end{equation}
and finally, the instanton path has to cross $x_{\cri}=0$ at $t=0$,
and then leaves the vicinity of $x_{\cri}$ with
\begin{equation}
\begin{casesb}
x_{\bl}^{(0)}(0) & =0\,,\\
x_{\bl}^{(0)}(t) & \underset{t\rightarrow+\infty}{\longrightarrow}+\infty\,.
\end{casesb}\label{eq:match4}
\end{equation}
To this order we find a solution which obeys the symmetry
\begin{equation}
\begin{casesb}
x_{\bl}^{(0)}(t) & \leftrightarrow-x_{\bl}^{(0)}(-t)\,,\\
v_{\bl}^{(2)}(t) & \leftrightarrow v_{\bl}^{(2)}(-t)\,,\\
P_{v,\bl}^{(0)}(t) & \leftrightarrow1-P_{v,\bl}^{(0)}(-t)\,,\\
P_{x,\bl}^{(0)}(t) & \leftrightarrow P_{x,\bl}^{(0)}(-t)\,.
\end{casesb}\label{eq:symmetries}
\end{equation}
The solution can be computed numerically using an iterative
procedure. Fig.~\ref{fig:solution} displays the resulting solutions.

The last step is to match the boundary layer solution at $t\rightarrow-\infty$
to the bulk solution in the limit $t\rightarrow0$, order by order
in $\tau$. In the bulk we have
\begin{align*}
v(t=\frac{t'}{\tau'^{2/3}}) & 
\underset{\phantom{\tau\rightarrow+\infty}}{=}
 \left(A_{0}+\frac{1}{\tau^{1/3}}A_{1}+\frac{1}{\tau^{2/3}}A_{2}\right)e^{t'/\tau'^{2/3}}+O(1/\tau)\\
 & \underset{\tau\rightarrow+\infty}{\sim}\left(A_{0}+\frac{A_{1}}{\tau^{1/3}}+\frac{A_{2}}{\tau^{2/3}}\right)\left(1+\frac{t'}{\tau'^{2/3}}\right)+O(1/\tau)
\,,
\end{align*}
and the boundary layer solution $v_{\bl}(t')$ up to second order gives
\begin{align*}
v_{\bl}(t') & 
\underset{\phantom{\tau\rightarrow+\infty}}{=}
\beta v_{\bl}^{(0)}(t)+\frac{1}{\tau'^{1/3}}\beta v_{\bl}^{(1)}(t')\frac{1}{\tau'^{2/3}}\beta v_{\bl}^{(2)}(t')+O(1/\tau')\\
 & \underset{t'\rightarrow-\infty}{\sim}\beta+\frac{1}{\tau'^{2/3}}\beta\left(C+t'\right)+O(1/\tau')\,,
\end{align*}
where $C>0$ is some positive constant (see Fig.~\ref{fig:solution}). Self consistency of the boundary
layer expansion and the bulk solution requires $A_{0}=\beta$, $A_{1}=0$, and
\begin{equation}
A_{2}=\frac{\beta C}{\left(\sqrt{\gamma\beta}\right)^{2/3}}\,.\label{eq:A2}
\end{equation}
$C$ can be expressed in term of the numerical solution of Eqs.~(\ref{eq:BL1-1}-\ref{eq:BL2-1}).
Using the equation for $v_{\bl}^{(2)}(t)$ we have
\[
\dot{v}_{\bl}^{(2)}-1=2\left(P_{v,\bl}^{(0)}-1\right).
\]
Both sides of the latter equation can be integrated, yielding
\[
\int_{-\infty}^{0}\frac{{\rm d}}{{\rm d}t}\left(v_{\bl}^{(2)}-t\right){\rm d}t=2\int_{-\infty}^{0}\left(P_{v,\bl}^{(0)}-1\right){\rm d}t\,.
\]
Using the symmetry~(\ref{eq:symmetries}) for $P_{v,\bl}^{(0)}$ and
the definition of $C$, we get
\begin{equation}
C=v_{\bl}^{(2)}(0)+2\int_{0}^{+\infty}P_{v,\bl}^{(0)}(t){\rm d}t,\label{eq:overshoot}
\end{equation}
which can be easily evaluated numerically to give $C\approx 0.8120$\,.

\medskip
\noindent{\bf The quasipotential for {\textnormal{$x>x_{\cri}$}}}: 
Taking into account the rescaling of time, the equations of motion and the fact that  the instanton path satisfies $H=0$, one finds after some algebra that the quasipotential  (See. Eq.~(\ref{eq:Hpath})) can be written as
\begin{equation}\label{eq:action2}
\Phi_{\tau}=\int_{-\infty}^{+\infty}\left(\tilde{P}_{v}(t)\right)^{2}{\rm d}t\,,
\end {equation}
where we denoted by $\tilde{P}_{v}(t)$ the instanton solution. The strategy goes as follows: we separate the contributions from the bulk and from the boundary layer in integral~(\ref{eq:action2}), and we evaluate both terms separately. In order to obtain a finite contribution for the boundary layer integral, we have to subtract the bulk solution from the boundary layer solution. This means that we first split the instanton path 
into two parts as
\[
\tilde{P}_{v}=P_{v}(t)+\left(\tilde{P}_{v}(t)-P_{v}(t)\right),
\]
where $P_{v}$ is the bulk solution, found in Sec.~\ref{subsec:Bulk-solution}.
We then write $\Phi_{\tau}$ as
\begin{align}
\Phi_{\tau} & =\int_{-\infty}^{+\infty}\left(P_{v}(t)+\left(\tilde{P}_{v}-P_{v}(t)\right)\right)^{2}{\rm d}t\nonumber \\
 & =\int_{-\infty}^{+\infty}\left(P_{v}(t)\right)^{2}{\rm d}t+2\int_{-\infty}^{+\infty}P_{v}(t)\left(\tilde{P}_{v}(t)-P_{v}(t)\right){\rm d}t+\int_{-\infty}^{+\infty}\left(\tilde{P}_{v}(t)-P_{v}(t)\right)^{2}{\rm d}t\,.\label{eq:decomposition}
\end{align}
Next we define $f_{\bl}=\tilde{P}_{v}-P_{v}$. By construction $f_{\bl}\rightarrow0$
outside the boundary layer and $f_{\bl}$ has the same scaling as the boundary
layer solution $P_{v,\bl}$. More precisely $f_{\bl}$ is  a natural function of $\tau'^{\,2/3}t$.
We now turn to compute the three different contributions in the right hand side of Eq.~(\ref{eq:decomposition}):

Using Eq.~(\ref{eq:solutionNth}), the bulk contribution is 
\begin{align}
\int_{-\infty}^{+\infty}\left(P_{v}(t)\right)^{2}{\rm d}t & =\int_{-\infty}^{0}\left(A_{0}+\frac{A_{2}}{\tau^{2/3}}\right)^{2}e^{2t}\,{\rm d}t\nonumber \\
 & =\frac{A_{0}^{2}}{2}+\frac{A_{0}A_{2}}{\tau^{2/3}}+O(\frac{1}{\tau})\label{eq:contribution1}
\end{align}
The bulk-boundary layer cross contribution can be equivalently written using the definition of $f_{\bl}$
\begin{align*}
2\int_{-\infty}^{+\infty}P_{v}(t)\left(\tilde{P}_{v}(t)-P_{v}(t)\right){\rm d}t & =2\int_{-\infty}^{+\infty}P_{v}(t)f(\tau'^{2/3}t)\,{\rm d}t\\
 & =\frac{2}{\tau'^{2/3}}\int_{-\infty}^{+\infty}P_{v}(t/\tau'^{2/3})f_{\bl}(t)\,{\rm d}t
\end{align*}
Using again Eq.~(\ref{eq:solutionNth}) and the expansion of $P_v$ and $f_{\bl}$ to zeroth order we get
\begin{align*}
2\int_{-\infty}^{+\infty}P_{v}(t)\left(\tilde{P}_{v}(t)-P_{v}(t)\right){\rm d}t  \underset{\tau\rightarrow+\infty}{\sim}\frac{2A_{0}}{\tau'^{2/3}}\int_{-\infty}^{0}f_{\bl}^{(0)}(t){\rm d}t+O(1/\tau),
\end{align*}
where 
\[
f_{\bl}^{(0)}=\beta\left(P_{v,\bl}^{(0)}(t)-\mathrm{1}_{]-\infty,0]}\right)
.
\]
Using then 
the symmetry of $P_{v,\bl}^{(0)}(t)$ (see Eq.~(\ref{eq:symmetries})), we get
\begin{equation}
2\int_{-\infty}^{+\infty}P_{v,\bu}(t)\left(\tilde{P}_{v}(t)-P_{v,\bu}(t)\right){\rm d}t\underset{\tau\rightarrow+\infty}{\sim}-\frac{2\beta A_{0}}{\tau'^{2/3}}\int_{0}^{+\infty}P_{v,\bl}^{(0)}(t){\rm d}t+O(1/\tau)\,.\label{eq:contribution2}
\end{equation}
The pure boundary layer contribution in Eq.~(\ref{eq:decomposition}) can be written
\begin{align}
\int_{-\infty}^{+\infty}\left(\tilde{P}_{v}(t)-P_{v}(t)\right)^{2}{\rm d}t & 
\underset{\phantom{\tau\rightarrow+\infty}}{=}
\int_{-\infty}^{+\infty}\left(f_{\bl}(\tau'^{2/3}t)\right)^{2}{\rm d}t\nonumber \\
 & \underset{\tau\rightarrow+\infty}{\sim}\frac{1}{\tau'^{2/3}}\int_{-\infty}^{+\infty}\left(f_{\bl}^{(0)}(t)\right)^{2}{\rm d}t+O(1/\tau)\nonumber \\
 & \underset{\tau\rightarrow+\infty}{\sim}\frac{2}{\tau'^{2/3}}\int_{0}^{+\infty}\left(P_{v,\bl}^{(0)}(t)\right)^{2}{\rm d}t+O(1/\tau)\label{eq:contribution3}
\,,
\end{align}
where we have again used in the last equality the symmetry of $P_{v,\bl}^{(0)}(t)$ (see Eq.~(\ref{eq:symmetries})).

Finally, summing the three contributions~(\ref{eq:contribution1}, \ref{eq:contribution2}, \ref{eq:contribution3})
and using the expressions for $A_{0}$ and $A_{2}$ in Eqs.~(\ref{eq:A2}-\ref{eq:overshoot}),
we arrive to our main result
\begin{equation}
\forall x>x_{\cri},\;\Phi_{\tau}(x)=\frac{\beta^{2}}{2}+\frac{\beta^{2}}{\left(\sqrt{\gamma\beta}\tau\right)^{2/3}}\left[v_{\bl}^{(2)}(0)+2\int_{0}^{+\infty}\left(P_{v,\bl}^{(0)}(t)\right)^{2}{\rm d}t\right]+O(1/\tau)\label{eq:main result}
\,.
\end{equation}
Moreover, the computation has shown that the bulk solution satisfies
$P_{v}(t)=0$ to any order in $\tau$, for $t>0$. This implies that
the part of the instanton path after $x_{\cri}$ is a relaxation path,
and has no contribution to the action~(\ref{eq:action2}). In particular, Eq.
(\ref{eq:main result}) is valid for any $x_{\cri}<x\leq x_{\sa}$ outside
a boundary layer of typical size $\frac{1}{\tau^{1/3}}\sqrt{\frac{\beta}{\gamma}}$
around $x_{\cri}$. 

\paragraph{Comments on the solution :}
\label{sec:comments}

\begin{enumerate}
\item Both the leading order and the first correction in the limit $\tau\rightarrow\infty$ are \emph{non-local}, that is, they only depend on the structure of $U(x)$ close to its inflection
point $x_{\cri}$, and not on the part $x>x_{\cri}$ Note that  if $U'''(x_{\cri})=0$, using similar
arguments as was done to find the scaling of the boundary layer, one
expects a correction scaling as $\frac{1}{\tau^{4/7}}$ instead of $\frac{1}{\tau^{2/3}}$.
\item The correction at order $\frac{1}{\tau^{2/3}}$ given by Eq.~(\ref{eq:main result})
has two origins. The first contribution comes from the fact that, for
finite $\tau$, the Ornstein--Uhlenbeck process has to overshoot the
velocity fluctuation above $\beta$. The fluctuation of the Ornstein--Uhlenbeck
process up to the value $\beta+\frac{\beta}{\tau'^{2/3}}v_{\bl}^{(2)}(0)$
is simply $\frac{\left(\beta+\frac{\beta}{\tau'^{2/3}}v_{\bl}^{(2)}(0)\right)^{2}}{2}=\frac{\beta^{2}}{2}+\frac{\beta^{2}}{\tau'^{2/3}}v_{\bl}^{(2)}(0)+O(1/\tau^{4/3})$.
The integral correction in Eq.~(\ref{eq:main result}) can be interpreted
as the price to pay for sustaining the active velocity at the value
$\beta$ during the crossing of the inflection point. It thus comes
from the fact that crossing the inflection point takes a finite time,
when $\tau$ is finite.
\item Whereas the correction at order $\frac{1}{\tau^{1/3}}$ vanishes,
there is no reason for the correction at order $\frac{1}{\tau}$ to
be zero, because the bulk instanton part has a contribution to this
order. One can thus expect the next order to scale as $\frac{1}{\tau}$
.
\item A numerical resolution of the boundary layer equations~(\ref{eq:BL1-1}-\ref{eq:BL2-1})
 gives $v_{\bl}^{(2)}(0)\approx0.5905$ and $2\int_{0}^{+\infty}\big(P_{v,\bl}^{(0)}(t)\big)^{2}{\rm d}t=0.2215$.
The sum of these two values gives the numerical constant $C$ of the result announced in Eq.~(\ref{eq:expansion large tau 2-1}).
For the escape problem ($x=x_{\cri}$), our result coincide with that of~\cite{bray_path_1990}, including for the numerical estimate of the constant $C$ (note also that the dominant order of~(\ref{eq:expansion large tau 2-1}) was derived earlier in~\cite{luciani_functional_1987}). 
In~\cite{bray_path_1990} the perturbation theory presents no boundary layer.

\end{enumerate}

\subsection{The quasipotential for \textnormal{$x < x_{\cri}$}}
We now turn to the computation of the quasipotential for $x<x_{\cri}$. In the following, we only sketch the computations as they are very similar to the case $x>x_{\cri}$. We look for an instanton path solving Eqs.~(\ref{eq:hamilton space2}-\ref{eq:hamilton momentum}) with the boundary conditions Eqs.~(\ref{eq:BC1}).
\subsubsection{Bulk solution}
The argument developed in Sec.~\ref{subsec:Bulk-solution} to show that $P_x=0$ to any order in $\tau$ remains valid in the range $x<x_{\cri}$. The instanton equations in
the bulk are
\begin{equation}
\begin{casesb}
\dot{x} & =\tau\left(v-U'(x)\right),\\
\dot{v} & =-v+2P_{v}\,,\\
\dot{P_{v}} & =P_{v}\,.
\end{casesb}\label{eq:bulk2}
\end{equation}
Contrary to the domain $x>x_{\cri}$, there is no anomalous scaling for the boundary layer in the part $x<x_{\cri}$.
We thus expand the solution as
\begin{equation}
P_{v}(t)=P_{v}^{(0)}(t)+\frac{1}{\tau}P_{v}^{(1)}(t)+\frac{1}{\tau^{2}}P_{v}^{(2)}(t)+...\;,
\label{eq:expansion bulk2}
\end{equation}
and similarly for $x(t)$ and $v(t)$. 
Thanks to the linearity of Eqs.~(\ref{eq:bulk2}) for $P_v$ and $v$, the solution to any order $k$ can be written
\begin{equation}
t<0:\begin{casesb}
v^{(k)}(t) & =A_{k}e^{t}\,,\\
P_{v}^{(k)}(t) & =A_{k}e^{t}\,.
\end{casesb}\label{eq:solutionNth2}
\end{equation}
One can then find the solution $x^{(k)}(t)$ using an expansion of $\dot{x}=\tau\left(v-U'(x)\right)$
to $k^{th}$ order in $\frac{1}{\tau}$. The constants $A_{k}$ have to be determined by matching the bulk solution to
the boundary layer as we do below to up to first order.
\subsubsection{Boundary-layer solution} 
The boundary layer equations are obtained from Eqs.~(\ref{eq:hamilton space2}-\ref{eq:hamilton momentum}) by rescaling time according to $t'=\tau t$ and expanding $U'$ in the vicinity of $x$ as
\begin{equation}
U'(x')=\beta+\gamma (x'-x)\,,
\end{equation}
where $\beta=U'(x)$ and $\gamma=U''(x)$. Note that both are strictly positive. We then define the new variable $y=x'-x$ and we get the boundary layer equations 
\begin{equation}
\begin{casesb}
\dot{y}_{\bl} & =v_{\bl} -\gamma y_{\bl}\,,\\
\dot{v}_{\bl}  & =-\frac{v_{\bl} }{\tau}+2\frac{P_{v,\bl} }{\tau}\,,
\end{casesb}\label{eq:hamilton space3}
\end{equation}
and 
\begin{equation}
\begin{casesb}
\dot{P_{x,\bl} } & =\gamma P_{x,\bl} \,,\\
\dot{P_{v,\bl} } & =\frac{P_{v,\bl} }{\tau}-P_{x,\bl} \,.
\end{casesb}\label{eq:hamilton momentum3}
\end{equation}

We then expand the boundary layer solution as
\[
y_{\bl}(t)=y_{\bl}^{(0)}(t)+\frac{1}{\tau}y_{\bl}^{(1)}(t)+\frac{1}{\tau^{2}}y_{\bl}^{(2)}(t)+...
\]
and use similar expansions for $P_{x,{\bl}},P_{v,{\bl}},v_{\bl}(t)$. We now turn to solve these equations order by order.

\paragraph{Zeroth order :} To this order we get
\begin{equation}\label{eq:zeroth2}
\begin{casesb}
y_{\bl}^{(0)}(t)&=0\\
v_{\bl}^{(0)}(t)&=\beta \\
P_{v,{\bl}}^{(0)}&=\beta(1-e^{\gamma t}) \;.
\end{casesb}
\end{equation}
Matching this with the bulk solution of Eq.~(\ref{eq:solutionNth2}) implies $A_0=\beta$.

\paragraph{First order :} To  order $1/\tau$ we get
\begin{equation}\label{eq:one2}
\begin{casesb}
v_{\bl}^{(1)}(t)&=\beta t +\frac{2\beta}{\gamma}(1-e^{\gamma t})\,,\\
P_{v,{\bl}}^{(1)}&=\beta t +\frac{2\beta}{\gamma}(1-e^{\gamma t})\,,
\end{casesb}
\end{equation}
and matching to the bulk solution of Eq.~(\ref{eq:solutionNth2}) implies $A_1=\frac{2\beta}{\gamma}$. Note that we don't need to derive an explicit expression for $y^{(1)}_{\bl}$ in order to calculate the quasipotential.

\paragraph{The quasipotential for $x<x_{\cri}$ :}
Using the same decomposition as in Eq.~(\ref{eq:action2}) and in the discussion that follows, we can express the quasipotential to order $1/\tau$ as
\begin{equation}
\Phi_{\tau}(x) \underset{\tau\rightarrow+\infty}{\sim}\int_{-\infty}^{0}\left(P_{v}^{(0)}(t)\right)^{2}{\rm d}t+\frac{2}{\tau}\int_{-\infty}^{0}P_{v}^{(0)}(t)P_{v}^{(1)}(t)\,{\rm d}t+\frac{2}{\tau}\int_{-\infty}^{0}P_{v}^{(0)}(t)f_{\bl}^{(0)}(t)\,{\rm d}t+\frac{1}{\tau}\int_{-\infty}^{0}\left(f_{\bl}^{(0)}(t)\right)^{2}{\rm d}t\,,\label{eq:decomposition2}
\end{equation}
where we have used again the notation $f_{\bl}=\tilde{P}_{v}-P_{v}$, with $\tilde{P}_{v}$ the instanton path and $P_v$ the bulk solution. A straightforward evaluation of the different terms in Eq.~(\ref{eq:decomposition2}) using the solutions Eqs.~(\ref{eq:solutionNth2}, \ref{eq:zeroth2}, \ref{eq:one2}) leads to the final result announced in Eq.~(\ref{eq:expansion large tau 1-1}), namely
\begin{equation}
\Phi_{\tau}(x) \underset{\tau\rightarrow+\infty}{\sim} \frac{\beta^2}{2}+\frac{\beta^2}{2\gamma\tau}+O(1/\tau^2)\,.
\end{equation}

\section{Perspectives}
\label{sec:perspectives}

We derived the escape rate and the stationary distribution of an AOUP, in expansions at small and large correlation time $\tau$.
By focusing on the asymptotics of small activity (in amplitude), we were able to use exact large-deviation techniques (\emph{i.e.}~without using the UCNA or the Fox approximation schemes that are often employed to study systems with colored noise).
In the $\tau\to 0$ limit the quasipotential is local to first order, as already known, but becomes non-local at second order in $\tau$. In the large $\tau$ limit the quasipotential is inherently non-local and singular.
This leads to a host of physical consequences such as a ratchet effect and a possible fly-over of metastable state facilitated by the memory of the active noise.

The results we have described open many questions.
The path-integral technique used enables one to derive the quasipotential as a functional of an optimal trajectory of an equilibrium problem;
as we focused, for simplicity, on the one-dimensional case, the explicit dependence on the time-dependent trajectory could be eliminated [see for instance the passage from~(\ref{eq:Phi2timedep}) to~(\ref{eq:2quasipot})].
This implies that the quasipotential can be reexpressed as a functional of the potential only and not of the optimal trajectory.
The generalization of our computation to higher dimensions is rather immediate, but it is not obvious that at the last stage one can similarly eliminate the explicit dependence on the optimal trajectory. 
This could potentially lead to interesting effects in cases where several optimal trajectories are in competition.

We have seen that the small-$\tau$ and the large-$\tau$ asymptotics present very different physical features:
at small $\tau$, the optimal trajectory remains close to the equilibrium one (that bypasses a potential barrier), while at large $\tau$ the dynamics is dominated by a ``barrier of force'' (\emph{i.e.}~the region of maximal force).
The switch from a potential to a force barrier dominated regime could either be a cross-over or signal a singularity as $\tau$ is increased. However, our perturbative approach leaves this question open.
Last, since our technique allows for a treatment of generic colored-noise dynamics (in the small noise limit), it could be instructive to compare its predictions to those of the UCNA and the Fox approximations in a systematic way.

\medskip

\noindent {\bf Acknowledgments:} 
We thank Nir Gov for his involvement in a project that led to this work,
and an anonymous referee for bringing our attention to Refs.~\cite{bray_instanton_1989,mckane_path_1990,bray_path_1990}.
EW and YK are supported by and ISF grant and an NSF-BSF grant. EW is partly supported by a Technion grant.
VL thanks the Technion, where part of this work was undergone, for kind hospitality.
VL is supported by the ERC Starting Grant No. 680275 MALIG, the ANR-18-CE30-0028-01 Grant LABS and the ANR-15-CE40-0020-03 Grant LSD.

\bibliographystyle{plain_url}
\bibliography{biblio}

\appendix

\section{Mean escape time from a metastable state\label{sec:Mean-escape-time}}

Let us now recall how the quasipotential is related to the mean escape
time from a metastable state $x_{0}$. Again in the following, $x_{\sa}$
is the saddle point to escape from the attraction basin of $x_{0}$.
We show how the mean escape time $\left\langle T_{\esc}\right\rangle $
can be computed directly from the quasipotential in the limit where
$D$ is small.

The standard procedure is to compute the mean escape time $\left\langle T_{\esc}\right\rangle $
from the Fokker--Planck equation with absorbing boundary
conditions: let $J(t)$ be the outgoing flux
at $x=x_{\sa}$, we have simply
\begin{equation}
\left\langle T_{\esc}\right\rangle =\int_{0}^{+\infty}tJ(t){\rm d}t.\label{eq:mean_esc_time}
\end{equation}
in general, there is no simple solution to this problem. But one can derive the result when escape is a rare event, which correspond to the small
$D$ limit in Eq.~(\ref{eq:AOUP model}). In this limit, the outgoing
flux is exponentially small in $1/D$, such that the relaxation time
$T_{\text{rel}}$ of the particle inside the trap is much smaller than the
mean escape time $\left\langle T_{\esc}\right\rangle $. A good approximation
is then to consider that two consecutive escapes are independent events,
which means that escape is a Poisson process with rate $\lambda$.
On the timescale $t\gg T_{\text{rel}}$, the probability $M(t)$ that the particle
is still in the trap within the interval $[0,t]$ follows the equation
\begin{equation*}
\frac{{\rm d}M}{{\rm d}t}=-\lambda M.
\end{equation*}
This relation can be immediately integrated to give 
\begin{equation*}
M(t)=e^{-\lambda t}.
\end{equation*}
 $J(t)$ is related to $M(t)$ through the simple relation
\begin{align}
J(t) & =-\frac{{\rm d}M}{{\rm d}t}
=\lambda e^{-\lambda t}.\label{eq:Jrate}
\end{align}
It follows from Eqs.~(\ref{eq:mean_esc_time}-\ref{eq:Jrate}) that
the mean escape time satisfies
\begin{equation}
\frac{1}{\left\langle T_{\esc}\right\rangle }=\lambda=\underset{T_{\text{rel}}\ll t\ll\left\langle T_{\esc}\right\rangle }{\lim}J(t).\label{eq:Jrelation1}
\end{equation}

On the other hand, $J$ is related to the transition probability through
\begin{equation}
J(t)=P\left(T_{\esc}=t\right)=P\left(x_{\sa},t|x_{0},0\right).\label{eq:Jrelation2}
\end{equation}
In the intermediate timescale regime $T_{\text{rel}}\ll t\ll\left\langle T_{\esc}\right\rangle $,
the transition probability $P\left(x_{\sa},t|x_{0},0\right)$ does not
depend on $t$ and is given by the large deviation principle
\begin{equation}
P\left(x_{\sa},t|x_{0},0\right)\underset{D\rightarrow0}{\asymp}e^{-\frac{\Phi_{\tau}(x_{\sa})}{D}}.\label{eq:Jrelation3}
\end{equation}
Relations~(\ref{eq:Jrelation1}-\ref{eq:Jrelation3})
together imply the large deviation result
\begin{equation}
\left\langle T_{\esc}\right\rangle \underset{D\rightarrow0}{\asymp}e^{\frac{\Phi_{\tau}(x_{\sa})}{D}}.\label{eq:asymp mean escape time appendix}
\end{equation}

\end{document}